\documentclass[a4paper,11pt]{article}
\usepackage{jheppub}
\usepackage[T1]{fontenc}
\usepackage{comment}
\usepackage[utf8]{inputenc}
\usepackage{amsmath}
\usepackage{slashed}
\usepackage{siunitx}
\usepackage{xcolor}
\usepackage[english]{babel}
\usepackage{subcaption}
\usepackage{indentfirst}
\usepackage[noabbrev]{cleveref}
\usepackage[force]{feynmp-auto} 
\usepackage{adjustbox}
\usepackage{graphicx}

\newcommand{\hc}{\mathrm{H.c.}}
\newcommand{\upq}{$\text{U}(1)_\text{PQ}$}
\newcommand{\fa}{$f_a$}
\newcommand{\gap}{\bar{g}_{a\gamma \gamma}}
\newcommand{\gaMM}{\bar{g}_{a N \gamma}}
\newcommand{\gae}{\bar{g}_{aee}}
\newcommand{\gaN}{\bar{g}_{aNN}}
\newcommand{\gamu}{\bar{g}_{a\mu\mu}}
\newcommand{\fsc}{\alpha_\text{em}}
\newcommand{\gapi}{\bar{g}_{a\pi\pi}}
\newcommand{\gall}{\bar{g}_{a\ell\ell}}

\newcommand{\Fp}{F_\pi}
\newcommand{\blue}[1]{\color{blue}{#1}}

\newcommand{\Ex}[1]{\cdot 10^{#1}}
\newcommand{\nn}{\nonumber \\}

\renewcommand{\O}{\mathcal{O}}
\newcommand{\op}[3]{\O^{#2,#3}_{#1}}

\title{\boldmath CP-violating axion interactions II: axions as Dark Matter}

\author[a,b]{V. Plakkot,}
\author[c]{W. Dekens,}
\author[a,b]{J. de Vries,}
\author[a,b]{and S. Shain}

\affiliation[a]{Institute for Theoretical Physics Amsterdam and Delta Institute for Theoretical Physics, University of Amsterdam, Science Park 904, 1098 XH Amsterdam, The Netherlands}
\affiliation[b]{Nikhef, Theory Group, Science Park 105, 1098 XG, Amsterdam, The Netherlands}
\affiliation[c]{Institute for Nuclear Theory, University of Washington, Seattle WA 91195-1550, USA}

\emailAdd{v.plakkot@uva.nl}
\emailAdd{wdekens@uw.edu}
\emailAdd{j.devries4@uva.nl}
\emailAdd{sshain@nikhef.nl }

\abstract{Axions provide a solution to the strong CP problem and are excellent dark matter candidates. The presence of additional sources of CP violation, for example to account for the matter/antimatter asymmetry of the universe, can lead to CP-violating interactions between axions and Standard Model fields. In case axions form a coherent dark matter background, this leads to time-oscillating fundamental constants such as the fine-structure constant and particle masses. In this work we compare the sensitivity of various searches for CP-odd axion interactions. These include fifth-force experiments, searches for time-oscillating constants induced by axion dark matter, and direct limits from electric dipole moment experiments. We show that searches for oscillating constants can outperform fifth-force experiments in the regime of small axion masses, but, in general, do not reach the sensitivity of electric dipole moment experiments. }

\begin{document}
	
\maketitle
\flushbottom

\section{Introduction}
The lack of CP violation in the strong sector via $\mathcal{L} \supset \bar{\theta} G \widetilde{G}$ (with $G$ being the gluonic field strength tensor, and $\widetilde{G}$ its dual)~\cite{Abel:2020pzs, PhysRevLett.116.161601,Dragos:2019oxn, Liang:2023jfj}, dubbed the strong CP problem, can be solved by allowing the parameter $\bar{\theta}$ to be a dynamical variable that settles to zero at the minimum of its potential. This is achieved by introducing a global U(1) Peccei-Quinn (PQ) symmetry, \upq, broken at a high energy scale \fa, called the axion decay constant~\cite{Peccei:1977hh,Peccei:1977ur}. The (pseudo-)Goldstone boson of this broken symmetry is the QCD axion, $a(x)$, and is an excellent dark matter (DM) candidate~\cite{Weinberg:1977ma,Wilczek:1977pj,Preskill:1982cy,Abbott:1982af,Dine:1982ah}. The so-called \emph{invisible} axion models offer attractive UV-complete mechanisms to introduce axions~\cite{Kim:1979if, Shifman:1979if, Zhitnitsky:1980tq, Dine:1981rt}. The model space for axions is however not limited to these two benchmark models; several extensions and alternative models provide ways to extend the viable parameter space of QCD axions~\cite{Kaplan:2015fuy, DiLuzio:2016sbl, Ballesteros:2016xej,  DiLuzio:2017pfr, DiLuzio:2020wdo, Plakkot:2021xyx, Berbig:2022pye, Diehl:2023uui}. Experimental efforts are yet to yield conclusive evidence for the existence of axions, although huge chunks of the available parameter space are yet to be probed~\cite{Irastorza:2018dyq,Sikivie:2020zpn}.

The PQ mechanism efficiently removes the single source of CP violation in the strong sector. However, even within the Standard Model (SM) this is not sufficient to remove all CP violation. In addition, it is not unlikely that additional sources of CP violation exist in a beyond-the-SM theory that addresses some of the SM shortcomings (such as the universal matter-antimatter asymmetry). The presence of CP-violating interactions beyond the QCD $\bar \theta$ term leads to CP-violating interactions between axions and SM fields. In recent work, we derived the CP-odd interactions between axions and leptons, hadrons, and nuclei in the framework of the Standard Model effective field theory (SM-EFT) \cite{Dekens:2022gha}. The CP-odd interactions lead to an axion-mediated scalar-scalar (monopole) and scalar-pseudoscalar (monopole-dipole) potential between, for example, atoms \cite{Moody:1984ba} that can be potentially detected in dedicated experiments, see e.g. Ref.~\cite{OHare:2020wah} for an overview. We concluded however that at least at the level of dimension-six SM-EFT interactions, it will be extremely difficult to detect axion CP-odd interactions because limits from electric dipole moments (EDMs) on the CP-odd dimension-six couplings are too stringent.

In this work, we will investigate whether we can detect CP-odd axion interactions under the assumption that axions form the dark matter (DM) in our universe. 
If we assume axions to form a coherent wave-like DM field new phenomenological implications arise. The axion DM field becomes locally coherent, with only a time-varying component, such that the equation of motion is solved by $a(t) \simeq a_0 \cos\left(m_a t\right)$, with $m_a$ the axion mass which thus sets the oscillation frequency of the axion DM wave. $a_0$ is related to the local axion density, $\rho_a \simeq \frac{1}{2}m_a^2 a_0^2$, which implies $a_0 = \sqrt{2 \rho_a}/m_a$~\cite{Marsh:2015xka}. Assuming then that axions account for all the DM, the axion density can be set to the local DM density, $\rho_a = \rho_\text{DM} \simeq 0.3 \text{ GeV/cm}^3$. Such a scenario can be tested by a range of axion experiments; see, for example Ref.~\cite{Irastorza:2018dyq} for an overview. In the presence of CP-violating axion interactions, the effects of the oscillating axion background field leads to time-varying fundamental constants such as the fine-structure constant as well as the masses of elementary particles and composite systems such as hadrons, and nuclei~\cite{Damour:2010rp,Graham:2011qk}. In this work, we extend the analysis of Ref.~\cite{Dekens:2022gha} to include the additional experimental signatures that arise for axionic DM. We will see that probes of variation in fundamental constants can be more sensitive than limits from fifth-force experiments, especially at small axion masses, but, unfortunately, in general still fall short of EDM experiments. 

This work is organized as follows. In Sect.~\ref{framework} we introduce the general setup, sources of CP violation, and effective CP-odd axion interactions and their renormalization. In Sect.~\ref{sec:expconstraints} we discuss experiments that are sensitive to axionic CP violation and compare constraints on different couplings from a broad range of experiments. We discuss specific beyond-the-Standard-Model scenarios in Sect.~\ref{sec:apps} and conclude in Sect.~\ref{discussion}.

\section{Theoretical framework and motivation}\label{framework}

Axions are interesting and well-motivated BSM particles that can account for several SM problems: the lack of a DM candidate, and a SM peculiarity, the CP-conserving nature of QCD. It must be said that within the SM small values of $\bar \theta$ are technically natural; once a small value of $\bar \theta \ll 1$ is selected it remains small. Only minuscule radiative corrections to $\bar \theta$ are induced \cite{Ellis:1978hq} which cannot be detected in present-day EDM experiments. 

\begin{figure}[t]
	\centering
	\includegraphics[width=0.95\linewidth, trim={3cm 0cm 14.7cm 2cm},clip]{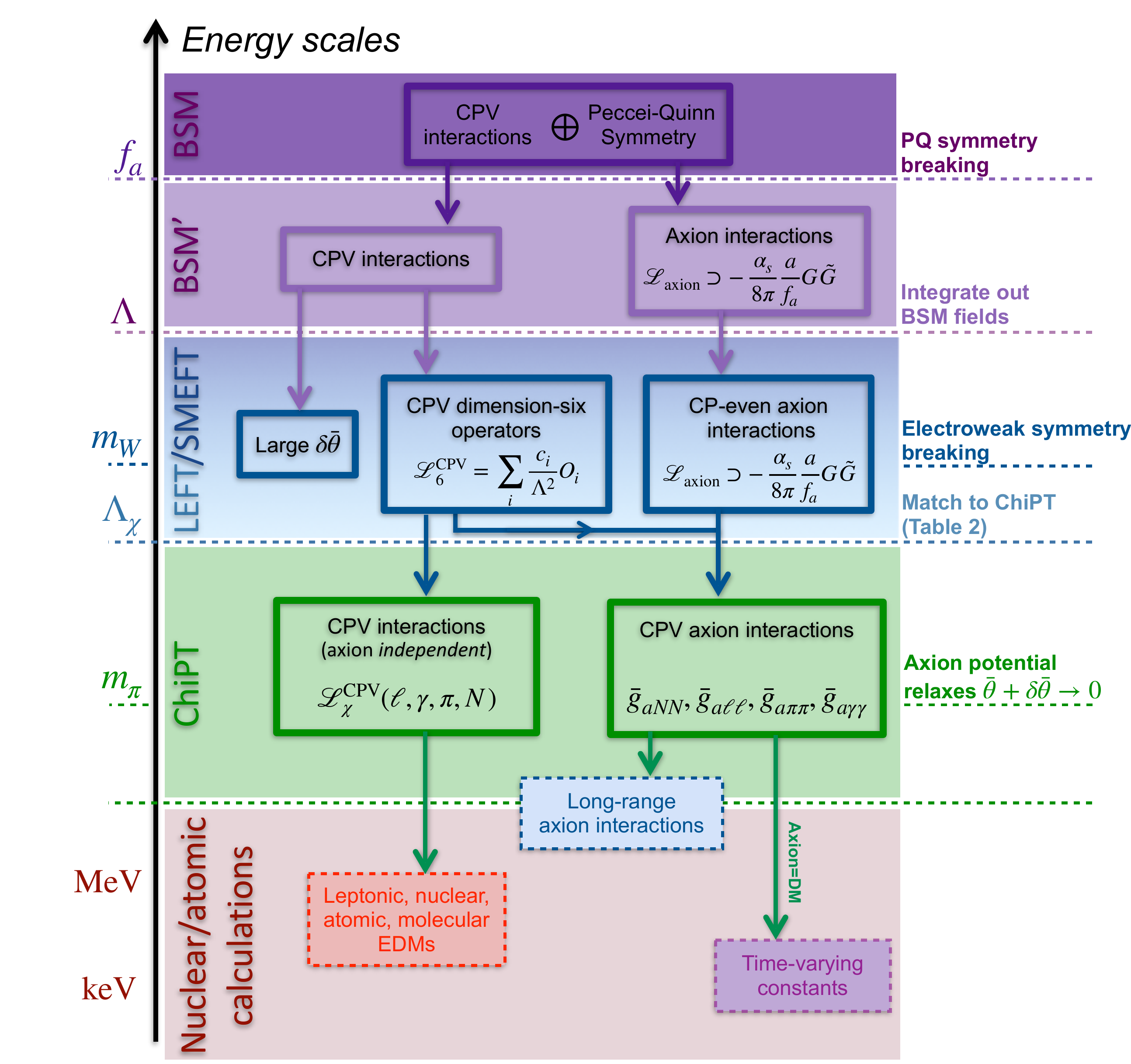}
	\caption{A schematic overview of the scales related to the PQ symmetry and CPV interactions that appear within the scenarios we consider in this work. The figure also shows how the new CPV interactions give rise to EDMs, while the interplay of these CPV sources with the PQ mechanism leads to CPV couplings of the axion to SM particles. These axion interactions can in turn be probed in fifth force experiments and, assuming the axion forms the DM abundance, searches for time-varying fundamental constants.}
	\label{fig:flowchart}
\end{figure}

In generic BSM extensions this is no longer the case. While one can still set $\bar \theta \ll 1$ at some energy scale, in general large corrections are induced at lower scales. For example, in supersymmetric models soft phases induce sizable threshold corrections \cite{Dine:2015jga} to $\bar \theta$. While models can be constructed that avoid tree- and one-loop level corrections to $\bar \theta$ \cite{Babu:1989rb}, higher loop corrections are still problematic \cite{deVries:2021pzl,Hisano:2023izx}. Similar conclusions can be drawn from a model-independent analysis within the SM-EFT framework  \cite{deVries:2018mgf}. The existence of dimension-six sources of CP violation imply large threshold corrections to $\bar \theta$ at the scale where the UV-complete theory is matched to the SM-EFT. This then strongly motivates an infrared mechanism to relax the value of $\bar \theta$ to zero, and the PQ mechanism is the only game in town. In essence, axions are even \textit{better} motivated in BSM scenarios than in the SM itself. We illustrate this in Fig.~\ref{fig:flowchart}.

In general, one may use the term ``axion'' to refer to a whole class of axion-like particles (ALPs) that arise similarly to the QCD axion, but need not solve the strong CP problem. Here, we restrict ourselves mainly to QCD axions, and the use of ``axion'' will mean QCD axion unless explicitly stated otherwise.

\subsection{CP-violating axion interactions in LEFT}
\label{sec:CPVaxint}
The Lagrangian describing the axion framework generally consists of the SM piece, the effective axion Lagrangian, and additional CP-violating sources,
\begin{align}
\mathcal L = \mathcal L_{\rm SM}+\mathcal L_{\rm axion}+\mathcal L_{\rm LEFT}\,.
\end{align}
The terms that depend on the axion are given by
\begin{align}
	\mathcal{L}_\text{axion} =\, &\frac{1}{2} (\partial_\mu a)^2 - \frac{\alpha_s}{8 \pi}\frac{a}{f_a}\widetilde{G}^A_{\mu \nu} G^{A \, \mu \nu} - \frac{1}{4}g_{a\gamma}
^{(0)} \frac{a}{f_a}\widetilde{F}_{\mu \nu} F^{\mu \nu} \nonumber \\
 &+ \sum_{f=\nu,e,q} \frac{\partial_\mu a}{2 f_a} \left[\bar{f}_L c_L^f \gamma^\mu f_L + \bar{f}_R c^f_R \gamma^\mu  f_R\right], 
 \label{eq:axlag}
\end{align}
where $f_a$ is the axion decay constant, $G_{\mu \nu}$ and $F_{\mu \nu}$ are the gluon and electromagnetic field strength tensors, $\alpha_s$ is the strong fine-structure constant, and $g_{a \gamma}^{(0)}$, $c_L^f$, and $c_L^f$ are coupling constants that are determined by the UV completion of the effective theory. The axion mass comes from low-energy non-perturbative QCD effects, and is related to the axion decay constant as~\cite{DiLuzio:2020wdo}
\begin{align}
	m_a \approx 6 \text{ $\mu$eV} \left(\frac{10^{12} \text{ GeV}}{f_a}\right).
\end{align}
The couplings of axions to SM particles are inversely proportional to $f_a$ and thus scale as $\propto m_a$. Nevertheless, the exact value of the couplings depend on the UV-complete model in question. Note however that this relation fails to hold for ALPs, where the couplings can be completely independent of $m_a$.

The Lagrangian \eqref{eq:axlag} can be extended with SM-EFT operators, which would be matched to the low-energy effective field theory (LEFT) at energies below electroweak scale. The contributions of such terms can be written as $\mathcal{L}_\text{LEFT} = \sum_i L_i \mathcal{O}_i$. Although the complete set of LEFT operators up to dimension six has been derived~\cite{Jenkins:2017jig}, we are only interested in the operators that induce CPV interactions. All such operators were considered in Ref.~\cite{Dekens:2022gha} and are listed in App.~\ref{app:BunchOfOps}, but we will only need a subset of operators here. All operators are suppressed by powers of the high-energy scale $\Lambda$, assumed to lie far above the electroweak scale, but well below the PQ scale $\sim f_a$.
	
\begin{figure}[t]
	\centering
	\includegraphics[width=0.75\linewidth]{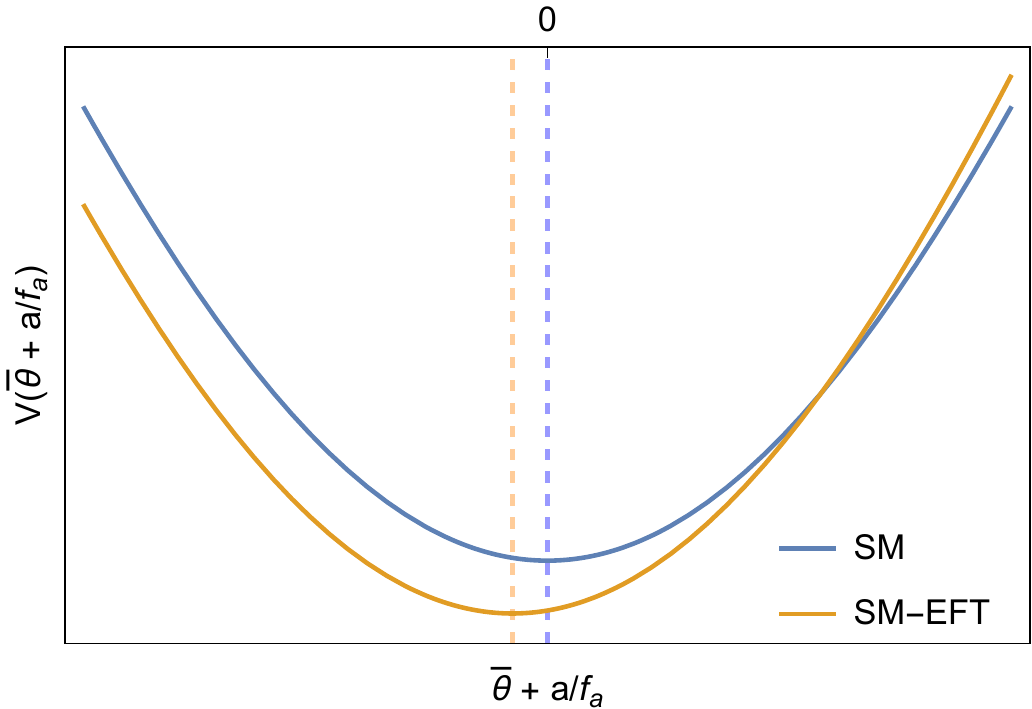}
	\caption{An (exaggerated) schematic representation of the shift in effective axion potential in the presence of higher-dimensional operators. The dashed lines represent the value of $\overline{\theta}+a/f_a$ at its potential minimum when SM-EFT operators are absent (in blue, coinciding with zero), and when they are present (in orange). The blue line corresponds to a scenario with just the $\bar{\theta}$ term and ignores CP violation in the electroweak interactions.}
	\label{fig:axpot}
\end{figure}

The PQ mechanism protects the naturalness of a small $\bar{\theta}$ by driving the axion potential to its minimum, which lies at $\bar{\theta} + \langle a \rangle /f_a = 0$. That is, the vacuum expectation value of the axion field (up to $f_a$) is such that it cancels out $\bar{\theta}$. In the presence of LEFT operators mentioned above, however, the minimum of the potential shifts such that $\bar{\theta} + \langle a \rangle/f_a \neq 0$. This is shown schematically in Fig.~\ref{fig:axpot}, where the blue (orange) line represents the potential without (with) LEFT operators. The field values at the minima are shown using dotted lines of the same color. 

The effective $\bar \theta$ no longer vanishes and some CP violation is left behind.  However, the remnant CP violation, characterised by the shift in the minimum, is at least suppressed by $\Lambda_\chi^2/\Lambda^{2}$ with $\Lambda_\chi\sim 1$ GeV, which (partially) explains its smallness. 

Because not all CP violation is removed we are led to consider CP-violating interactions between axions and SM fields.  
For experimental purposes we are mainly interested in CP-odd interactions of axions with leptons and hadrons. We derived these in detail in Ref.~\cite{Dekens:2022gha} and here we present the most relevant couplings for the present work. In particular, we consider
\begin{align}\label{axionCPV}
	\mathcal{L}_{\mathrm{axion},\,\mathrm{CPV}} =\,& \bar{g}^{(0)}_{a\ell\ell} \,a\, \bar{\ell} \ell + \gapi^{(0)}\, a\, \vec \pi\cdot\vec \pi+ \bar g^{(0)}_{a\pi_0\pi_0}\, a\, \pi_0\pi_0+ \bar{g}^{(0)}_{aNN} \,a\, \bar{N} N \nonumber\\
	&+  \bar{g}^{(0)}_{a\ell\gamma} \,a\, \bar{\ell} \sigma^{\mu \nu} \ell F_{\mu \nu} +  \bar{g}^{(0)}_{an \gamma} \,a\, \bar{n} \sigma^{\mu \nu} n F_{\mu \nu}  +  \bar{g}^{(0)}_{ap \gamma} \,a\, \bar{p} \sigma^{\mu \nu} p F_{\mu \nu}  \nonumber\\
	 & +  \frac{\gap^{(0)}}{4} \,a\, F_{\mu\nu} F^{\mu \nu}+ \ldots\,,
\end{align}
in terms of leptons $\ell = \{e,\mu,\tau\}$, the pion triplet $\pi^a$, the nucleon doublet $N=(p,\,n)^T$, and the photon field strength $F_{\mu\nu}$. The terms in the first line describe, respectively, CP-violating axion-lepton, isospin-conserving and -breaking axion-pion, and axion-nucleon interactions (we have omitted here a possible isospin-breaking axion-nucleon term as it only plays a small role in our analysis). The second line describes axion couplings to lepton and nucleon magnetic moments, while the last line describes the CP-odd counterpart of the usual axion-photon-photon interaction. We use a bar notation on the coupling constants to indicate that the coupling breaks CP and the $(0)$ superscript indicates that these are bare couplings in the Lagrangian which are renormalized at the loop level. We discuss how these couplings are generated as well as their renormalization in more detail below. 

\subsubsection*{ Standard Model contributions}
Although small, the SM does provide a source of CP violation beyond the $\bar \theta$ term in the form of the phase in the CKM matrix. The basis-independent quantity that induces CP violation is the Jarlskog invariant, $J = {\rm Im}(V_{ud}^*V_{us}V_{td}V_{ts}^*)\simeq 3\cdot 10^{-5}$, which can induce the couplings of Eq.\ \eqref{axionCPV}. The generated CP-odd axion couplings are expected to be proportional to $G_F^2 J$, as at least two exchanges of the $W$ boson are required to obtain the combination of CKM elements in $J$. To estimate the contributions to Eq.\ \eqref{axionCPV}, we start from an effective Lagrangian at a scale of $\mu \simeq 2$ GeV, where all heavy SM fields have been integrated out. The relevant sources of CP violation at this scale are then captured by dimension-six operators, which arise from diagrams involving the $W$ boson.

For the couplings to nucleons \cite{Georgi:1986kr,Okawa:2021fto} and leptons \cite{Dekens:2022gha} we use the estimates in the literature 
\begin{align}
\bar{g}_{a\ell\ell} &\sim \frac{m_\ell}{f_a}\left( \frac{\alpha}{4\pi}\right)^2\left(G_F F_\pi^2\right)^2 J\simeq \frac{m_\ell}{m_e} \frac{1}{2}\frac{10^{-25}\,{\rm MeV}}{f_a}\,,\\
\bar{g}_{aNN} &\sim \frac{m_*}{f_a}\left(G_F F_\pi^2\right)^2 J\simeq \frac{1}{2}\frac{10^{-18}\,{\rm MeV}}{f_a}\,.
\end{align}
To estimate the coupling to pions we consider the contribution coming from a tree-level weak operator $\sim V_{ud}V_{us}^*G_F(\bar u_L \gamma_\mu d_L) (\bar s_L \gamma^\mu u_L)$, combined with an electroweak penguin operator  $\sim\frac{\alpha}{4\pi} V_{td}V_{ts}^*G_F(\bar q_R \gamma_\mu Q q_R) (\bar s_L \gamma^\mu d_L)$, where $Q$ denotes the quark charge matrix.\footnote{One could in principle consider contributions from other tree-level interactions instead of an EW penguin operator. However, the chiral properties of the tree-level interactions only allow them to contribute to chirally suppressed operators involving additional derivatives in the chiral Lagrangian. The chiral suppression is expected to be similar to the loop suppression encountered when employing the EW penguin operators. } Using Naive Dimensional Analysis (NDA) \cite{Manohar:1983md,Gavela:2016bzc}, we obtain  
\begin{align}
\gapi &\sim \frac{F_\pi^2}{ f_a}\left(G_F F_\pi^2\right)^2 J\simeq 3 F_\pi \frac{10^{-17}\,{\rm MeV}}{f_a}\,.
\end{align}

Finally, we consider contributions to $\gap$ arising from $a-K$ mixing, induced by a tree-level weak operator, combined with a $K\gamma\gamma$ vertex induced by another tree-level weak operator and two insertions of the QED current. The corresponding NDA estimate is
\begin{align}
\gap &\sim \frac{\alpha}{4\pi} \frac{ 1}{f_a}\left(G_F F_\pi^2\right)^2 J\simeq 2\cdot 10^{-22}\frac{1}{ f_a}\,.
\label{eq:SMaxph}
\end{align}

When discussing the sensitivity of current and future experiments in Sect.~\ref{sec:expconstraints}, we will use the above estimates to indicate the expected SM effect, although in most cases they are too small to be shown.

\subsubsection*{ Higher-dimensional contributions}
The contributions to the couplings in the chiral Lagrangian of Eq.~\eqref{axionCPV} in terms of the quark-level LEFT operators is determined by hadronic matrix elements, parameterized by so-called low-energy constants (LECs).
The values of $\bar{g}^{(0)}_{a\ell\ell}$, $\gapi^{(0)}$, and $\bar{g}^{(0)}_{aNN}$ in terms of CP-violating dimension-six LEFT operators have been discussed in great detail in Ref.~\cite{Dekens:2022gha}, while our discussion of $\gap$ and the coupling of the axion to the magnetic moments go beyond that work.  Here we will not perform a detailed construction of the chiral Lagrangian that leads to the couplings in Eq.\ \eqref{axionCPV}. Instead, we summarize both the LEFT operators as well as the NDA estimates of their contributions to Eq.~\eqref{axionCPV} in App.\ \ref{app:BunchOfOps}. These estimates will be of use when we consider several specific BSM scenarios in Sect.~\ref{sec:apps}.
\\\\

The interactions in Eq.~\eqref{axionCPV} have a different form to those in the usual CP-conserving axion Lagrangian in which the axion couples derivatively  to electrons (see Eq.~\eqref{eq:axlag}), nucleons, and pions~\cite{DiLuzio:2020wdo}. As we discuss in more detail in the next section, this feature will lead to new, time-varying, effects.

\section{Time-varying effects}
In this section, we will consider the possible signals that arise when we assume that the axion explains the DM abundance. This assumption leads to a coherent background axion field. In the presence of such a field, the CPV couplings   effectively introduce a time-oscillating component to the lepton, pion, and nucleon masses \cite{Damour:2010rp}. For example, the effective electron mass becomes
\begin{align}\label{eq:me}
	m_e(t) \simeq m_{e} \left[1-\frac{\bar{g}_{aee}}{m_{e}} a(t)\right],
\end{align}
where $m_{e}$ denotes the time-independent electron mass term that originates in the SM. Similarly, the second and third line of Eq.~\eqref{axionCPV} lead to time-varying lepton and nucleon magnetic moments and a time-varying fine-structure constant. 

Without knowing the mechanism of CP violation it is not possible to address the relative sizes of the couplings in Eq.~\eqref{axionCPV}. However, all terms are expected
to be suppressed as $\bar g_i \propto \kappa^{d_i}\frac{\kappa^3}{f_a\Lambda}$, where $d_i$ is the dimension of the coupling $\bar g_i$ and $\kappa$ is a hadronic scale, $\kappa\lesssim \Lambda_\chi$. We will discuss more specific scenarios where this question can be answered in Sect.~\ref{sec:apps}. Before doing so, we will first discuss loop corrections that relate the various couplings in Eq.~\eqref{axionCPV} as well as experimental constraints, irrespective of the origin of the couplings. This is important because the experimental constraints on the various terms in Eq.~\eqref{axionCPV} are quite different. Roughly, the strongest direct limits are set on the axion-photon-photon coupling, while weaker constraints are set on the axion-nucleon-nucleon, magnetic axion-nucleon-nucleon, and axion-electron-electron interactions. Essentially no direct constraints are set on the axion-pion-pion terms but because they renormalize other interactions, they will still play an important role.

We stress that the kind of interactions discussed here are not unique to axions, and are usually studied in the context of ultralight scalar DM (ULDM) (see, e.g., Ref.~\cite{Damour:2010rp} where the protagonist is a dilaton). As a result, the constraints discussed in the next section, in the context of an effective field theory framework, apply equally well to ULDM. Although we will refer only to axions in the upcoming sections, the axionic nature of DM is employed only in Sect.~\ref{sec:apps} where we discuss specific BSM models.

\subsection{Axion-nucleon couplings and time-varying nucleon masses}

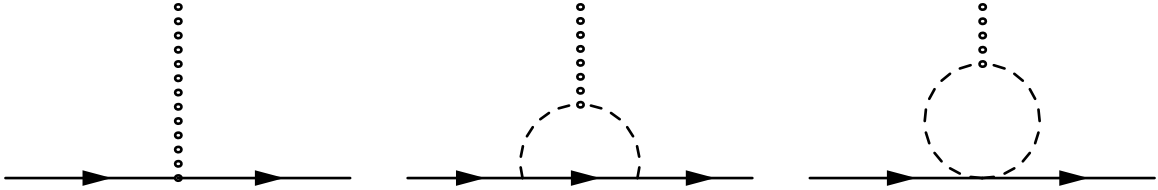
\begin{figure}[t]
    \centering
    \begin{minipage}{.3\textwidth}
    \centering
        \begin{fmffile}{axion_nucleon_a}
        \begin{fmfgraph*}(.3,.15) 
        \fmfstraight
        \fmfleft{i1,i2} 
        \fmfright{o1,o2}
        \fmftop{t1}
        \fmfbottom{b1}
        \fmf{fermion}{i1,b1,o1}
        \fmf{dbl_dots,tension=0}{t1,b1}
        \end{fmfgraph*}
        \end{fmffile}
    \end{minipage}
    \hfill
    \begin{minipage}{.3\textwidth}
    \centering
        \begin{fmffile}{axion_nucleon_b}
        \begin{fmfgraph*}(.3,.15) 
            \fmfstraight
            \fmfleft{i1,i2} 
            \fmfright{o1,o2}
            \fmftop{t1,t2,t3}
            \fmfbottom{b1,b2,b3,b4}
            \fmf{fermion}{i1,b2,b3,o1}
            \fmf{dbl_dots,tension=1.5}{t2,v}
            \fmf{dashes,left=0.5}{b2,v}
            \fmf{dashes, right=0.5}{b3,v}
        \end{fmfgraph*}
        \end{fmffile}
    \end{minipage}
    \hfill
    \begin{minipage}{.3\textwidth}
    \centering
        \begin{fmffile}{axion_nucleon_c}
        \begin{fmfgraph*}(.3,.15) 
        \fmfstraight
        \fmfleft{i1,i2} 
        \fmfright{o1,o2}
        \fmftop{t1}
        \fmfbottom{b1}
        \fmf{fermion}{i1,b1,o1}
        \fmf{dbl_dots,tension=4}{t1,v}
        \fmf{dashes,left}{b1,v}
        \fmf{dashes, right}{b1,v}
        \end{fmfgraph*}
        \end{fmffile}
    \end{minipage}
    \caption{Contributions to the axion-nucleon coupling. The solid, dashed, and dotted lines represent nucleons, pions, and axions, respectively.}
 	\label{fig:aNN}
\end{figure}

The CP-odd axion-nucleon interaction arises from the diagrams in Fig.~\ref{fig:aNN}. In principle there appear additional diagrams from electromagnetic corrections but these are not relevant for our analysis. The total effective interaction becomes
\begin{align}\label{axionCPVnucleon}
\bar{g}_{aNN} = \bar{g}^{(0)}_{aNN} + \frac{9 \pi }{2} \frac{m_\pi\,g_A^2}{(4 \pi \Fp)^2}\left(\gapi +\frac{1}{3}\bar{g}_{a\pi_0\pi_0}^{(0)}\right)\,,
\end{align}
in terms of the nucleon axial charge, $g_A \simeq 1.27$, the pion decay constant, $F_\pi \simeq 92.2$ MeV, and the pion mass, $m_\pi$. In the presence of an axion DM background, this terms lead to an oscillating nucleon mass
\begin{align}
	m_N(t) &= m_{N} \left[1-\frac{\bar{g}_{aNN}}{m_{N}} a(t)\right]\,. 
	\label{eq:mNvar}
\end{align}

\subsection{Axion-photon coupling and a time-varying fine-structure constant}
Conventional axion models lead to CP-even axion-photon interaction of the form
\begin{align}
	\mathcal{L}_{a \gamma}^{\text{CP}}  \supset g_{a\gamma\gamma} a F_{\mu \nu}\widetilde{F}^{\mu \nu},
	\label{eq:convaxph}
\end{align}
where the axion-photon coupling $g_{a\gamma \gamma}$ dictates the probability of axion-photon conversion in the presence of magnetic fields, as is probed in several axion detection experiments such as haloscopes and helioscopes (see, e.g., Refs.~\cite{PhysRevLett.51.1415, ADMX:2018gho, CAST:2013bqn}). CPV axion interactions can induce an additional effective axion-photon term to the Lagrangian,
\begin{align}
	\mathcal{L}_{a \gamma}^{\text{CPV}} = \frac{\gap}{4}a F_{\mu \nu} F^{\mu \nu}\,,
	\label{eq:CPVaxph}
\end{align}
which causes an interaction $\propto a(\vec E^{\,2}-\vec B^{\,2})$ in terms of electric and magnetic fields $\vec E$ and $\vec B$, as opposed to the conventional CP-even coupling $\propto a(\vec E \cdot \vec B)$.

\begin{figure}[t]
    \centering
    \begin{minipage}{.3\textwidth}
    \centering
    \begin{fmffile}{axion_photon_a}
    \begin{fmfgraph*}(.3,.15) 
        \fmfstraight
        \fmfleft{i1,i2,i3} 
        \fmfright{o1,o2,o3}
        \fmf{dbl_dots,tension=2}{i2,v1}
        \fmf{dashes,left=1}{v1,v2}
        \fmf{dashes,right=1}{v1,v2}
        \fmf{photon}{v2,o1}
        \fmf{photon}{v2,o3}
    \end{fmfgraph*}
    \end{fmffile}
    \end{minipage}
    \hfill
    \begin{minipage}{.3\textwidth}
    \centering
    \begin{fmffile}{axion_photon_b}
    \begin{fmfgraph*}(.3,.15) 
    \fmfstraight
    \fmfleft{i1,i2,i3} 
    \fmfright{o1,o2,o3}
    \fmftop{t1}
    \fmfbottom{b1}
    \fmf{dbl_dots,tension=2}{i2,v1}
    \fmf{dashes}{v1,t1}
    \fmf{dashes}{v1,b1}
    \fmf{dashes}{t1,b1}
    \fmf{photon}{b1,o1}
    \fmf{photon}{t1,o3}
    \end{fmfgraph*}
    \end{fmffile}
    \end{minipage}
    \hfill
    \begin{minipage}{.3\textwidth}
    \centering
    \begin{fmffile}{axion_photon_c}
    \begin{fmfgraph*}(.3,.15) 
    \fmfstraight
    \fmfleft{i1,i2,i3} 
    \fmfright{o1,o2,o3}
    \fmftop{t1}
    \fmfbottom{b1}
    \fmf{dbl_dots,tension=2}{i2,v1}
    \fmf{dashes}{v1,t1}
    \fmf{dashes}{v1,b1}
    \fmf{dashes}{t1,b1}
    \fmf{photon}{t1,o1}
    \fmf{photon}{b1,o3}
    \end{fmfgraph*}
    \end{fmffile}
    \end{minipage}
    \caption{One-loop contributions to the $a\gamma\gamma$ vertex. The dashed lines represent pions or leptons, while dotted and wavy lines denote axions and photons, respectively.}
 	\label{fig:feyn}
\end{figure}
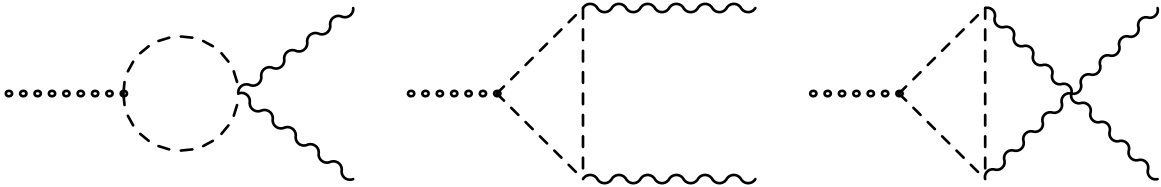

In addition to the direct pieces the CPV coupling gets contributions from the loop diagrams in Fig.~\ref{fig:feyn} involving virtual mesons and leptons\footnote{One could in principle think about diagrams with virtual protons appearing in the loop. However, these do not appear within a heavy-baryon $\chi$EFT framework where anti-nucleons are integrated out at low energies. In this language, such contributions are already contained in the direct piece $\gap^{(0)}$.}. The total coupling then becomes
\begin{align}\label{gap_renorm}
	\gap & = \gap^{(0)}-\frac{\fsc}{12\pi} \frac{\gapi}{m_\pi^2} -\sum_{e,\mu,\tau}\frac{2\fsc}{3 \pi}\frac{\bar{g}_{a\ell\ell}}{m_\ell},
\end{align}
where $\alpha_{\rm em}$ is the fine-structure constant. The constants $\alpha_{\rm em}$, $m_\pi$, and $m_\ell$ in the above expressions can be taken as time-independent as any time-varying component will appear at higher order in the already suppressed CPV axion couplings, and can be neglected. 

In the presence of a coherent DM axion background field, a CPV axion-photon term may be interpreted as an effective time variation in the fine structure constant. Combined with the kinetic term for the photons, 
\begin{align}
	\mathcal{L} \supset -\frac{1}{4}F^{\mu \nu} F_{\mu \nu} + \frac{\gap}{4} a F^{\mu \nu} F_{\mu \nu}.
	\label{eq:kinplusaxphot}
\end{align}
Following Ref.~\cite{Damour:2010rp}, a field redefinition $A_\mu \rightarrow A_\mu/e$ ensures that the factors of $e$ (and thus the fine structure constant) appear only in the terms given in Eq.~\eqref{eq:kinplusaxphot}. The two terms  can then be combined by defining an effective fine-structure constant, 
\begin{align}
	\mathcal{L} \supset -\frac{1}{4e^2}(1-\gap)F^{\mu \nu} F_{\mu \nu} \equiv -\frac{1}{16\pi\alpha_{\rm eff}}F^{\mu \nu} F_{\mu \nu},
\end{align}
which can be physically interpreted as a time-oscillation in the fine-structure constant,
\begin{align}
\fsc(t) &\simeq \fsc{} \left[1+\gap \, a(t)\right] \,.
\label{eq:fscvar}
\end{align}

\subsection{Magnetic axion couplings and the time-varying magnetic moments}
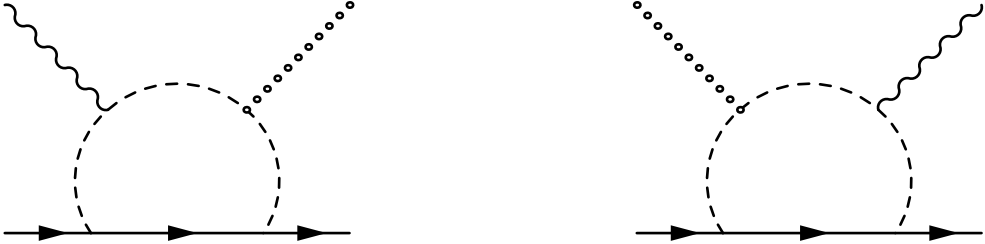
\begin{figure}[t]
    \centering
    \begin{minipage}{.45\textwidth}
    	\centering
    \begin{fmffile}{mag_mom_a}
    \begin{fmfgraph*}(.3,.2) 
    \fmfstraight
    \fmfleft{i1,i2,i3} 
    \fmfright{o1,o2,o3}
    \fmftopn{t}{7}
    \fmfbottomn{b}{5}
    \fmf{fermion}{i1,b[2]}
    \fmf{fermion}{b[2],b[4]}
    \fmf{fermion}{b[4],o1}
    \fmf{dashes,left=.38,tension=.85}{b[2],v1,v2,b[4]}
    \fmf{photon}{v1,t[1]}
    \fmf{dbl_dots}{v2,t[7]}
    \end{fmfgraph*}
    \end{fmffile}
    \end{minipage}
    \hfill
    \begin{minipage}{.45\textwidth}
    	\centering
    \begin{fmffile}{mag_mom_b}
    \begin{fmfgraph*}(.3,.2) 
    \fmfstraight
    \fmfleft{i1,i2,i3} 
    \fmfright{o1,o2,o3}
    \fmftopn{t}{7}
    \fmfbottomn{b}{5}
    \fmf{fermion}{i1,b[2]}
    \fmf{fermion}{b[2],b[4]}
    \fmf{fermion}{b[4],o1}
    \fmf{dashes,left=.38,tension=.85}{b[2],v1,v2,b[4]}
    \fmf{dbl_dots}{v1,t[1]}
    \fmf{photon}{v2,t[7]}
    \end{fmfgraph*}
    \end{fmffile}
    \end{minipage}
    \caption{Corrections to nucleon magnetic moments via the axion-pion portal. The solid, dashed, dotted, and wavy lines represents nucleons, pions, axions, and photons, respectively.}
	\label{fig:axMM}
\end{figure}

The pionic terms in Eq.~\eqref{axionCPV} induce axion-nucleon magnetic couplings at the one-loop level. The only non-vanishing diagrams are depicted in Fig.~\ref{fig:axMM}, and they induce
 \begin{align}\label{axionmag}
\bar{g}_{ap \gamma}  & = \bar{g}^{(0)}_{ap \gamma}  - \frac{e \pi \bar{g}_{a\pi\pi}}{2 m_\pi}\frac{g_A^2}{(4 \pi \Fp)^2}\,,\nonumber\\
\bar{g}_{an \gamma}  & = \bar{g}^{(0)}_{an \gamma}  + \frac{e \pi \bar{g}_{a\pi\pi}}{2 m_\pi}\frac{g_A^2}{(4 \pi \Fp)^2}\,.
\end{align}
Together, these terms can be interpreted as time-varying nucleon magnetic moments 
\begin{align}
	\mu_p(t) &\simeq \mu_{p} \left[1 + \left(\bar{g}^{(0)}_{ap \gamma}
	- \frac{e \pi \bar{g}_{a\pi\pi}}{2 m_\pi}\frac{g_A^2}{(4 \pi \Fp)^2}\right)\frac{a(t)}{\mu_{p}}\right]\,,\nonumber \\
\mu_n(t) &\simeq \mu_{n} \left[1 + \left(\bar{g}^{(0)}_{an\gamma}
+ \frac{e \pi \bar{g}_{a\pi\pi}}{2 m_\pi}\frac{g_A^2}{(4 \pi \Fp)^2}\right)\frac{a(t)}{\mu_{n}}\right]\,,	
	\label{eq:magmom}
\end{align}
in terms of the proton $\mu_p \simeq 2.79\,\mu_N$ and neutron $\mu_n \simeq -1.93\,\mu_N$ magnetic moments in units of nuclear magnetons. While a nuclear magneton $\mu_N = e/(2 m_p)$, the explicit dependence on the proton mass is conventional and there is no direct link between the oscillating nucleon mass and the oscillating nucleon magnetic moment. Any such dependence is hidden in the direct pieces which cannot be computed in a model-independent way. Nevertheless, we expect that the two matrix elements are connected and one may estimate the time-dependence in the direct piece to come from the proportionality $g_{aN\gamma}^{(0)} \propto 1/m_N(t)$ given in Eq.~\eqref{eq:mNvar}. In this way we can estimate the sensitivity of time-varying nucleon masses through their impact on the time-varying nucleon magnetic moment. 

For elementary particles such as leptons the link between masses and magnetic moments is cleaner. In this case, the leptonic magnetic moments obtain time-dependence through the $a\bar \ell \ell$ couplings and we obtain 
\begin{align}
	\mu_\ell(t) &\simeq \mu_\ell \left[ 1 + 2\left(\bar{g}^{(0)}_{a\ell \gamma}  
	+ \bar g_{a\ell \ell} \frac{\mu_{\ell}}{m_\ell}\right)\frac{a(t)}{\mu_{\ell}}\right]\,,
\end{align}
where $\mu_\ell = e/2 m_{\ell}$. 

\section{Constraining the CPV axion interactions}
\label{sec:expconstraints}
 
\subsection{Atomic clocks and interferometers}
The discussion in Sect.~\ref{sec:CPVaxint} shows that a coherent light axion DM field with CPV interactions can have effects similar to ULDM~\cite{Damour:2010rp}, which include a time-variation in fundamental constants such as particle masses, the fine-structure constant, and magnetic moment of nucleons. Several of these effects can be probed with precision experiments using atomic clocks~\cite{Arvanitaki:2014faa}. Atomic transition frequencies depend on the nuclear magnetic  moment $\mu_A$, the fine-structure constant, and electron mass through
\begin{align}
	f_A(t) \propto \left[ \frac{\mu_A(t)}{\mu_e (t)}\right]^{\zeta_A} \left[\fsc(t)\right]^{\xi_A + 2} \left[m_e(t)\right],
	\label{eq:fA_em}
\end{align}
where $\mu_e(t)$ is the electron magnetic moment, $\zeta_A$ is either 1 or 0 depending on whether the transition is hyperfine or optical. $\xi_A$ depends on the atomic properties and is an effect of relativistic and many-body corrections which are independent of $m_e$~\cite{Dzuba:1998au}; see, e.g., Ref.~\cite{Arvanitaki:2014faa} for $\xi_A$ values of some relevant systems. We consider atoms with a single valence nucleon in the core implying that the nuclear magnetic moment can be written as 
$\mu_A(t) \simeq \mu_N(t) + \Delta(t)$, where $\mu_N$ is the magnetic moment of the valence nucleon (see Eq.~\eqref{eq:magmom}) and $\Delta(t)$ denotes nuclear contributions. The latter are hard to compute from first principles. We expect that the dominant time-dependence arises from the axion-pion interactions which modifies the pion mass and thus the nucleon-nucleon potential. Barring an exact understanding of these contributions, we will focus on the contributions from $\mu_N(t)$ instead. 

Under these assumptions, we obtain\footnote{The dependence on $\bar g_{aN\gamma}/\mu_A$ is somewhat different with respect to Ref.~\cite{Flambaum:2004tm}, which instead finds $\bar g_{an\gamma}/\mu_n$ for a system with a valence neutron and thus neglects nuclear contributions to the magnetic moments which are not negligible. } 
\begin{align}
	\left(\frac{\mu_A(t)}{\mu_e(t)}\right) &\simeq \left(\frac{\mu_A}{\mu_e}\right) \left[ 1 + \left(\frac{\bar{g}_{aN\gamma}}{\mu_{A}}  -\frac{\bar{g}^{(0)}_{a e \gamma}}{\mu_e} - \frac{\bar{g}_{aee}}{m_{e}} \right) a(t) \right]\,.
\end{align}
Consequently, for the ratio of transition frequencies of two atoms $A$ and $B$, we get a fractional change
\begin{align}
	\frac{\delta (f_A/f_B)}{(f_A/f_B)} \approx \left[\left( \frac{\zeta_{A}}{\mu_A}-\frac{\zeta_{B}}{\mu_B}\right) \bar{g}_{aN\gamma} -\zeta_{AB}\left(\frac{\bar{g}^{(0)}_{a e \gamma}}{\mu_e}+\frac{\bar{g}_{aee}}{m_e}\right) + \xi_{AB} \,\gap\right] a(t),
	\label{eq:freqvar}
\end{align}
where $\zeta_{AB} = \zeta_A - \zeta_B$ and $\xi_{AB} =\xi_A - \xi_B$. Factors of $\left[\fsc(t)\right]^2$ and the direct dependence on $m_e(t)$ cancel in the ratio. With this expression we can relate frequency stability of a system of atomic clocks to the strengths of different CPV axion interactions. Take, for example, a model where only the last term in the square brackets contributes. Knowing $\xi_{AB}$ and $\rho_a$ (which enters through $a_0$), a non-observation of oscillation in the frequency ratio lets us draw ($m_a$-dependent) limits on $\gap$, up to the inherent frequency stability of the systems $A$ and $B$ which appears on the left-hand side of Eq.~\eqref{eq:freqvar}. 

Atom interferometry-based gravitational wave detectors have also been proposed as candidates for ULDM detection~\cite{Arvanitaki:2016fyj}. The proposed setup relies on differential phase accumulation between spatially separated atom interferometers, of the same type of atoms in this case, using controlled laser pulses. Since only one atomic system is used, it is clear from Eq.~\eqref{eq:fA_em} that the dependence of variation in transition frequency on $m_e(t)$ is explicit and the Rydberg factor of 2 in the exponent of $\fsc(t)$ also comes into play. For electronic transitions (i.e., with $\zeta_{A}=0$), the transition frequency then varies as
\begin{align}
	\frac{\delta f_A}{f_A} \simeq \left[(\xi_A + 2)\,\gap +\frac{\bar{g}_{aee}}{m_e}\right] a(t),
	\label{eq:atintvar}
\end{align}
and thus is sensitive to CPV axion-photon and axion-electron interactions. Although such systems have not been realised yet, we use the predicted reach of the proposals to draw projected limits on various CPV axion couplings~\cite{Arvanitaki:2016fyj,AEDGE:2019nxb,Badurina:2019hst}.

\subsection{Fifth-force experiments}
In the presence of above-mentioned CPV axion interactions, long-distance axion exchange may lead to a new gravitation-like effective force, dubbed the ``fifth force'', thereby violating Newton's inverse-square law (ISL) and the weak equivalence principle (WEP). Experiments looking for such fifth forces are powerful tools to constrain CPV axion-nucleon and axion-lepton couplings \cite{Berge:2017ovy,Touboul:2017grn,Wagner:2012ui,Hees:2018fpg}. Moreover, given the Coulombic contribution to the nucleon binding energy, the constraints also help us draw conservative limits on $\gap$ as well~\cite{Damour:2010rp}. In what follows, we will be heavily using the fifth-force search results to interpret bounds on different CPV axion interactions; we refer the reader to Ref.~\cite{Dekens:2022gha} for more details.

\subsection{Dipole interaction searches}
CPV interactions in the effective theory, involving axions or otherwise, can induce EDMs in various systems such as nuclei, atoms, and molecules. Similarly, the presence of both CP-odd and CP-even operators involving axions can lead to spin-dependent monopole-dipole interactions. For the CP-even couplings, we use the DFSZ model and set the vacuum expectation values of the two scalar doublets to be equal. Without going into the details (once again, the interested reader is directed to Ref.~\cite{Dekens:2022gha}), we would just like to point out that EDM measurements~\cite{Abel:2020pzs,PhysRevLett.116.161601,Bishof:2016uqx,ACME:2018yjb,Roussy:2022cmp,Ema:2021jds} set very stringent bounds on specific BSM scenarios, while proposed monopole-dipole searches such as ARIADNE and QUAX~\cite{ARIADNE:2017tdd,ARIADNE:2020wwm,Crescini:2016lwj,Crescini:2017uxs} can be competitive with EDM searches for specific axion mass ranges. We do not show the resulting constraints in the general framework discussed in this section, but rather employ them in Sect.~\ref{sec:apps}.

\subsection{Limits on CPV axion couplings}
\paragraph{The CP-violating axion-photon coupling.} The effect of axion DM on $\fsc$ can be isolated by focusing on optical transitions only, so that $\zeta_{A} = \zeta_{B} =0$. In this way, we can focus on $\gap$ exclusively. Assuming a coherent wave-like background axion field, a suitable approximation for cold and feebly-interacting axion DM with small $m_a$ and homogeneous local distribution, the experimental sensitivity depends on the fractional stability of the atomic clocks, the (axion) DM density ($\rho_\text{DM}$), the axion mass, as well as the coherence time of the DM field. 
The scaling of sensitivity with axion mass however implies that they remain competitive only for tiny $m_a$.  For higher masses, the sensitivity decreases rapidly making such probes unviable at large axion masses. In this regime individual (averaged) measurements are not able to record the variation in $\fsc$ given the high frequency of oscillation (recall that the frequency of a coherent axion DM field is given by $m_a$, and 1 eV $\simeq 1.52 \times 10^{15}$ Hz) and the limits to which individual measurement times can be discretised, and we rely on alternative methods to limit $\gap$.

For very light axions, roughly $m_a \leq 10^{-10}$ eV,  limits on the time-oscillation in $\fsc$ provide very stringent upper bounds on $\gap$. Given the rapid progress in this field, more stringent bounds are expected in the future as clocks with even better frequency stability are designed. Atom interferometry can also become very sensitive for ultralight DM axion in the future~\cite{Arvanitaki:2016fyj, AEDGE:2019nxb, Badurina:2019hst}. For larger axion masses, fifth-force experiments probing long-range axion-mediated forces, effectively violating the weak equivalence principle (WEP), provide upper bounds on $\gap$~\cite{Berge:2017ovy, Wagner:2012ui, Hees:2018fpg}. Laser-interferometric methods~\cite{Stadnik:2014tta,Grote:2019uvn,Vermeulen:2021epa}, as well as the gravitational wave detector AURIGA~\cite{Branca:2016rez}, have recently also been shown to be competitive with these bounds near $m_a=10^{-12}$ eV. Above $\sim 10 \, \mu\text{eV}$, the constraints instead arise from the bounds on axion-to-photon conversion given by various conventional axion experiments and astrophysical bounds on $g_{a \gamma \gamma}$, since the axion-to-photon conversion does not discriminate between the two operators in Eqs.~\eqref{eq:convaxph} and \eqref{eq:CPVaxph}. In general, we can expect $\gap$ to be smaller than its CP-conserving counterpart and that the effect of $\gap$ on such experiments be unnoticeable compared to a possibly larger $g_{a\gamma\gamma}$. However, this is a model-dependent statement and here we just show the bounds on $\gap$. 

\begin{figure}[t!]
	\centering
	\includegraphics[width=0.9\linewidth]{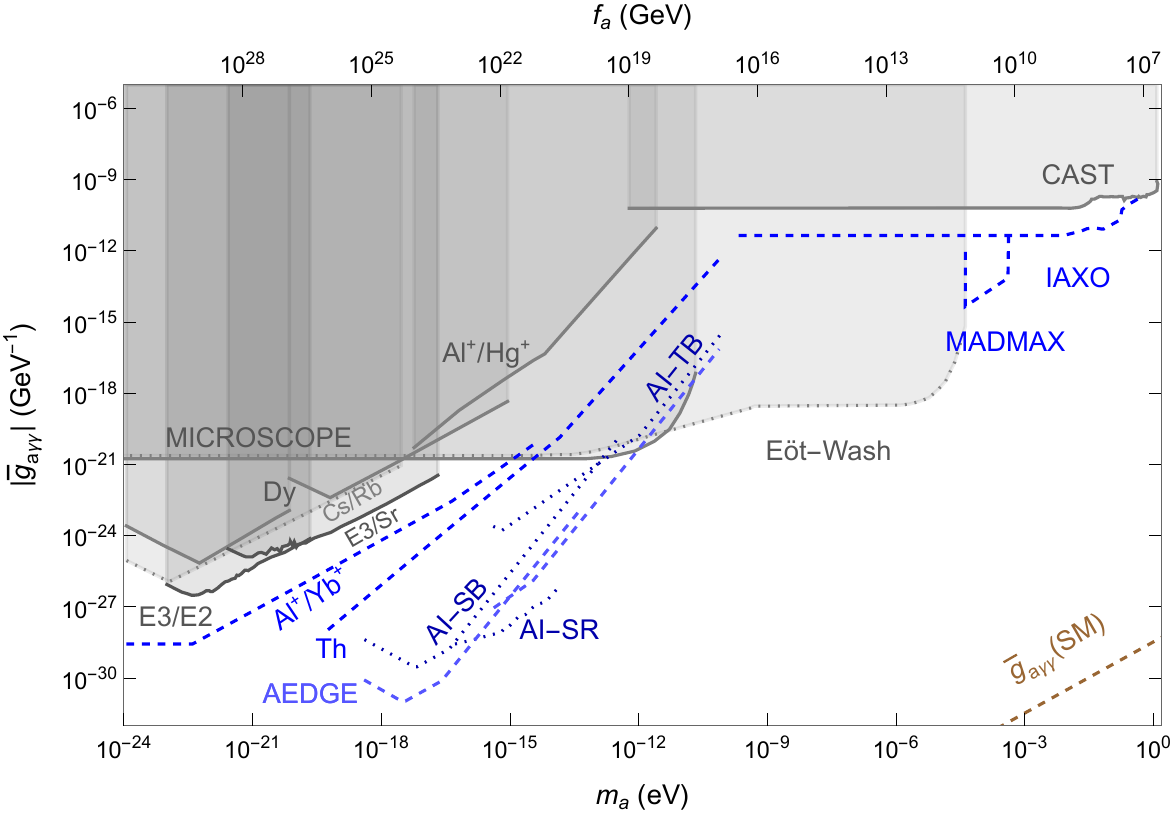}
	\caption{Upper bounds on the CPV axion-photon coupling $\gap$. The existing (projected) bounds are given in gray (blue). Atomic clock (labeled Dy, Cs/Rb, Al$^+$/Hg$^+$,  Al$^+$/Yb$^+$, E3/E2, E3/Sr), nuclear clock (Th), and atom interferometer (AI-TB, AI-SR, AI-SB, AEDGE, AION-km) bounds rely on the CPV structure of the axion-photon coupling to produce a time-oscillating $\fsc$~\cite{VanTilburg:2015oza,Hees:2016gop,Arvanitaki:2014faa,Arvanitaki:2016fyj,Kalaydzhyan:2017jtv, Campbell:2012zzb, AEDGE:2019nxb, Badurina:2019hst, Filzinger:2023zrs}. CAST, MADMAX, IAXO are conventional axion experiments that look for axion-photon conversion in the presence of magnetic fields~\cite{Sokolov:2022fvs, CAST:2013bqn, Li:2021oV, Caldwell:2016dcw, IAXO:2020wwp}. MICROSCOPE and E\"ot-Wash probe look for WEP violation through axion-mediated long-range forces~\cite{Berge:2017ovy,Touboul:2017grn,Wagner:2012ui,Hees:2018fpg}. Several other (weaker) bounds are not shown here to avoid clutter. The SM contribution via Eq.~\eqref{eq:SMaxph} ($\gap$(SM)) is shown in brown.}
	\label{fig:axphoton}
\end{figure}

Fig.~\ref{fig:axphoton} shows the current (in grey) and projected (in blue) bounds on $\gap$.  For $m_a \lesssim 10^{-17}\text{ eV}$, current atomic clock experiments (labeled by Dy, Cs/Rb, E3/E2 and E3/Sr)~\cite{Hees:2016gop, VanTilburg:2015oza, Filzinger:2023zrs}) already set constraints on $\gap$ that are up to six orders of magnitude more stringent than fifth-force searches (MICROSCOPE).  A recent analysis of the Sr/Cs system~\cite{Sherrill:2023zah} also improves on the WEP bounds but is weaker than the limits from E3/E2 and E3/Sr, and the limits are not shown here to avoid clutter. Bounds from laser interferometry and AURIGA around $m_a = 10^{-12}$ eV~\cite{Stadnik:2014tta,Grote:2019uvn,Branca:2016rez,Vermeulen:2021epa} are also not shown for the same reason.

Future atomic clocks (labeled by Al$^+$/Yb$^+$) can become more sensitive by several orders of magnitude and increase the axion mass range~\cite{Arvanitaki:2014faa}. Nuclear clocks, labeled by Th, are very promising as well~\cite{Campbell:2012zzb}. AI-SB, AI-SR, AI-TB refer to atom interferometry projections for terrestrial and space-based gravitational wave detectors, and AEDGE, AION-km label the projected reach for cold Strontium atom interferometers~\cite{Arvanitaki:2016fyj,AEDGE:2019nxb,Badurina:2019hst}. Such experiments can improve the constraints by many orders of magnitude.

\begin{figure}[t!]
		\centering
		\includegraphics[width=0.9\linewidth]{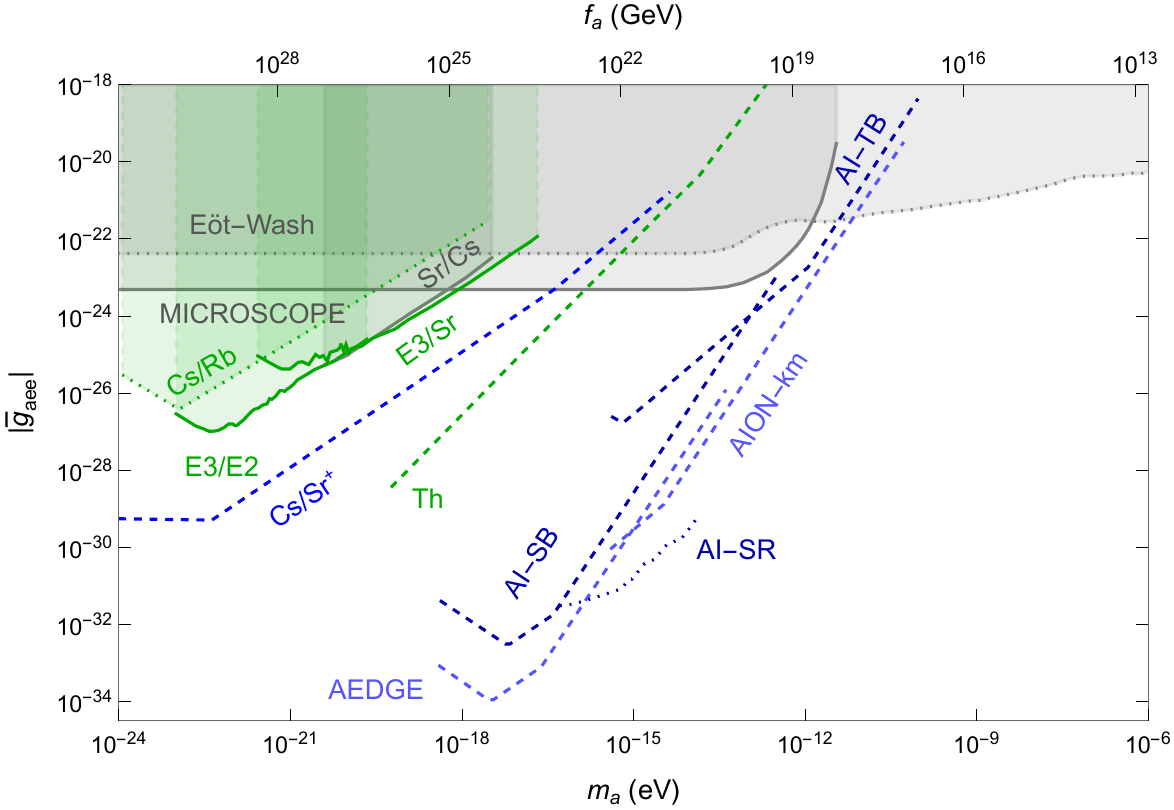}
		\caption{Direct limits (in grey, shaded) and projections (in blue) on $\gae$. The WEP bounds are from Ref.~\cite{AxionLimits}, and Sr/Cs is from Ref.~\cite{Sherrill:2023zah}. Cs/Sr$^+$ is from Ref.~\cite{Arvanitaki:2014faa}; see Fig.~\ref{fig:axphoton} for the rest of the projections. In green are indirect bounds on $\gae$ obtained using Eq.~\eqref{gap_renorm}, where shading implies existing limits~\cite{Hees:2016gop, Filzinger:2023zrs}. The SM effect discussed previously is too small ($\sim 10^{-41}$ at $m_a= 10^{-6}$ eV) to enter the frame.}
		\label{fig:gaemodind}
\end{figure}

\paragraph{The CP-violating axion-electron coupling.} Atomic clock systems involving hyperfine transitions (e.g., with microwave clocks) are sensitive to small variations in $\mu_A/\mu_e$, and thus to $\gaMM$, $\bar g^{(0)}_{a\ell\gamma}$, and $\gae$ through Eq.~\eqref{eq:freqvar}.  The limits and projections on $\gae$ from microwave clock measurements as well as other experiments are shown in Fig.~\ref{fig:gaemodind}. 

The gray lines labelled by MICROSCOPE and E{\"o}t-Wash indicate again current constraints from fifth-force experiments. The only limit on $\gae$ from microwave clock systems which improve upon these WEP limits come from a recent analysis of Sr/Cs clocks~\cite{Sherrill:2023zah} which is also depicted in gray (and labeled with Sr/Cs). It is also possible to extract upper limits on axion-lepton couplings (as well as $\gapi$) with the use of only optical systems through Eq.~\eqref{gap_renorm}, assuming there are no large cancellations against the others terms appearing in that equation. Despite the loop suppression, because of the superior sensitivity of optical clocks to a time-varying fine-structure constant, the resulting indirect limits on $\gae$ are similar to the more direct limits from hyperfine transitions. These indirect bounds are in shown in green in Fig.~\ref{fig:gaemodind}. Once again, we do not show the recent constraints close to $m_a = 10^{-12}$ eV~\cite{Stadnik:2014tta,Grote:2019uvn,Branca:2016rez,Vermeulen:2021epa} that are competitive with fifth-force probes to avoid clutter.

Looking ahead, atom interferometers are sensitive to variations in the electron mass and $\fsc$ and promise a direct reach in  $\bar{g}_{aee}$ (see Eq.~\eqref{eq:atintvar}) that is about two orders of magnitude more sensitive than the translated bounds from $\gap$ (using Eq.~\eqref{gap_renorm})~\cite{AEDGE:2019nxb, Badurina:2019hst, Arvanitaki:2016fyj}. 

\begin{figure}[t!]
	\centering
	\includegraphics[width=0.9\linewidth]{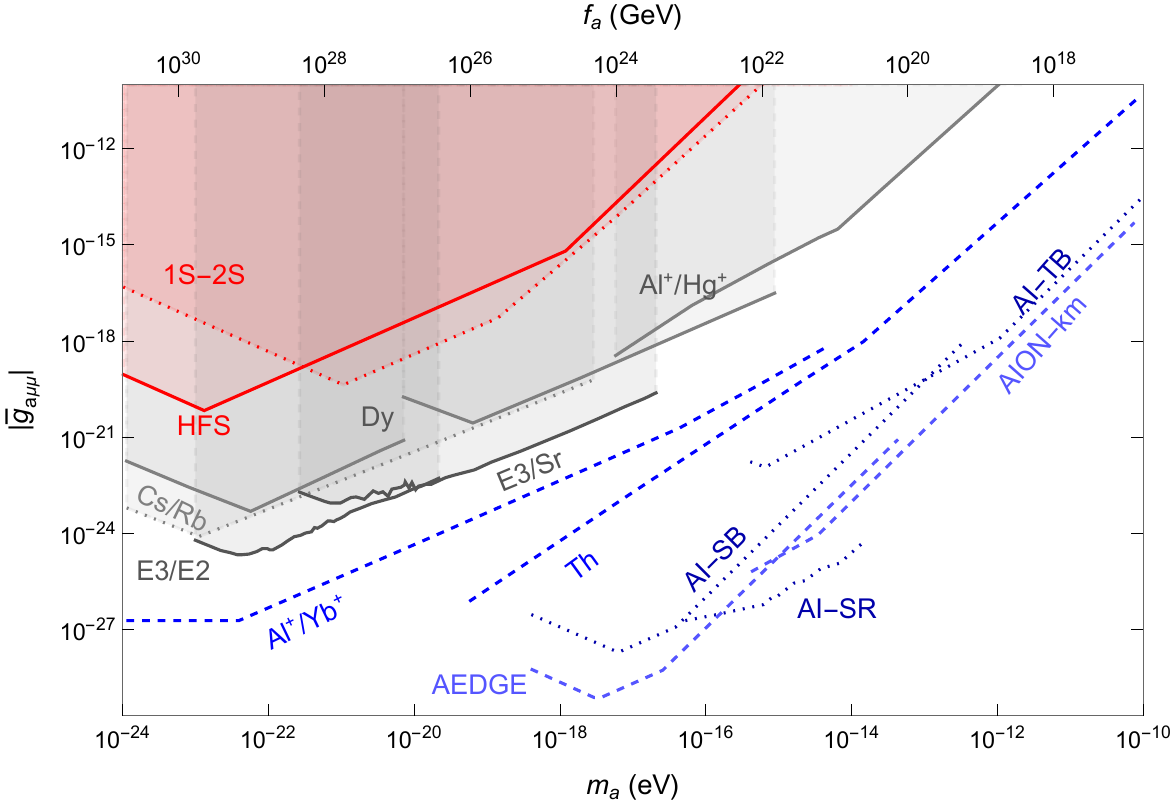}
	\caption{Bounds on the CPV axion-muon coupling. The bounds in red are from a recent study on muonium and muonic atoms, probing the variation in muon mass with hyperfine splitting (HFS) and 1S--2S transitions~\cite{Stadnik:2022gaa}. In gray and blue are the current and projected upper bounds coming from the loop-induced axion-photon coupling, given in Fig.~\ref{fig:axphoton}. The WEP bounds (not shown here) become stronger than the best projection at $m_a \sim 10^{-12}\text{ eV}$ (similar to Fig.~\ref{fig:axphoton}). The SM effect is $\sim 10^{-43}$ at $m_a=10^{-10}$ eV.}
	\label{fig:axmu}
\end{figure}

\paragraph{The CP-violating axion-muon coupling.} We now briefly discuss the constraints on a coupling that does not enter Eq.~\eqref{eq:freqvar} explicitly, namely $\gamu$. 
A recent study pointed out that $\gamu$ can be studied by its effect on the spectra of muonium and muonic atoms~\cite{Stadnik:2022gaa} which would provide direct bounds.  However, as shown in Fig.~\ref{fig:axmu}, these direct bounds pale in comparison to the upper bounds put on $\gamu$ by indirect constraints through Eq.~\eqref{gap_renorm}, as is confirmed in Ref.~\cite{Batell:2021ofv}. There are in principle other limits, e.g., through muon precession experiments~\cite{Janish:2020knz}, that are weaker than the muonium limits and are thus not shown in Fig.~\ref{fig:axmu}.

\begin{figure}[t!]
	\centering
	\includegraphics[width=0.9\linewidth]{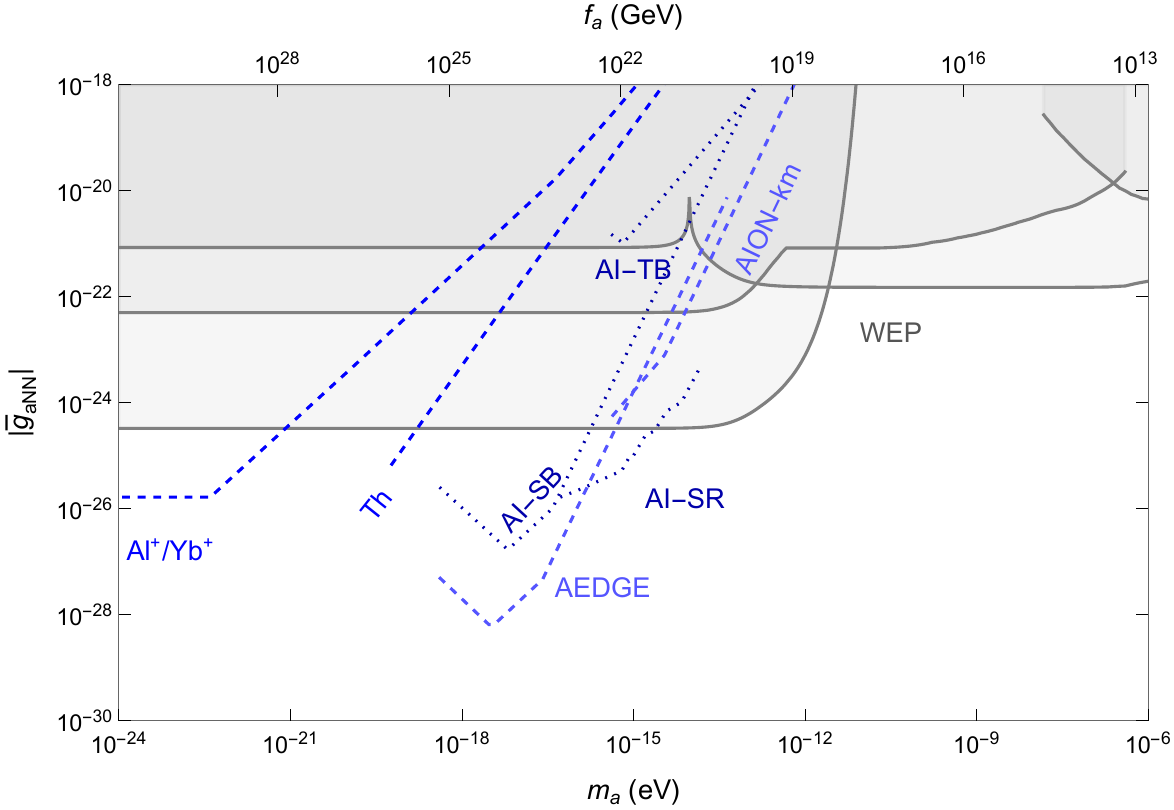}
	\caption{Limits and projections on $\gaN$. See Fig.~\ref{fig:axphoton} for details on the projections, and Refs.~\cite{Dekens:2022gha, OHare:2020wah} for the WEP bounds. The SM effect is once again small, $\sim 10^{-34}$ at $m_a=10^{-6}$ eV.}
	\label{fig:gaNmodind}
\end{figure}

\paragraph{The CP-violating axion-nucleon coupling.} The axion-nucleon coupling $\gaN$ plays a role through its effects on $\gaMM$ due to variation in nucleon mass over time, see the discussion below Eq.~\eqref{eq:magmom}. In principle, there could be similar effects on $\bar g_{a\gamma\gamma}$. However, there are no closed nucleon loops in heavy-baryon $\chi$EFT, such that $\gaN$ cannot contribute through the first diagram in Fig.\ \ref{fig:feyn}. Instead,
such high-energy effects are captured in the LEC $\bar g_{a\gamma\gamma}$ itself, implying that knowledge of the quark-level theory is needed in order to relate $\bar g_{a\gamma\gamma}$ and  $\gaN$. 
Keeping this in mind, we close our eyes and estimate the effects of $\gaN$ on $\gap$ by using Eq.~\eqref{gap_renorm} as we did for $\gall$.  The resulting contributions are $m_N$-suppressed and give weaker bounds, which are shown in Fig.~\ref{fig:gaNmodind}. As mentioned before, if the direct contribution of axion DM to nucleon magnetic moments is assumed to scale as $1/m_N(t)$, we can also estimate the resulting limits on the axion-nucleon coupling coming from hyperfine transitions. The best estimated projection from a Cs/Sr$^+$ system~\cite{Arvanitaki:2014faa} lies close to the Al$^+$/Yb$^+$ line and is not shown here. Unlike the previously discussed interactions, no existing data can improve the WEP bounds (and are thus not shown in Fig.~\ref{fig:gaNmodind}), and only proposed experiments do better. 
Again we stress, that the limits are not computed in a consistent way and we will provide more accurate computations in specific high-energy scenarios in the section below.

\paragraph{Axion-pion coupling.} Pion loops play a part in all of the above-mentioned observables, and thus $\gapi$ is in principle detectable from several terms in Eq.~\eqref{eq:freqvar}. Nevertheless, one can expect the contribution to $\gap$ to dominate in terms of constraining power due to larger scale suppressions in other terms, as well as better frequency stability of optical clocks. In the absence of other direct measurements in the variation of pion mass, we rely completely on the bounds arising from Eq.~\eqref{gap_renorm}. The resulting constraints would look similar to Fig.~\ref{fig:axphoton} with the vertical axis rescaled by about one order of magnitude. As we will see below in various BSM scenarios, the pion couplings play an important role. 

\section{The bigger picture }
\label{sec:apps}

So far our analysis has been performed at the level of Eq.~\eqref{axionCPV} without specifying the sources of CP violation beyond the $\bar \theta$ term. As such, the results in Figs.~\ref{fig:axphoton}-\ref{fig:gaNmodind} do not depend on the axionic nature of Dark Matter, making them equally valid for ULDM. In the rest of this work, we will consider several BSM scenarios and study how the CP-violating terms in Eq.~\eqref{axionCPV} are generated. We subsequently discuss the sensitivity of searches for oscillating constants due to axionic DM and compare them to EDM searches that constrain additional sources of CP violation directly. We again borrow heavily from our older work~\cite{Dekens:2022gha} where we have provided the necessary EDM computations and here we focus on the results. We will study three different scenarios:
\begin{enumerate}
\item An explicit BSM scenario involving scalar leptoquarks.
\item A minimal left-right symmetric scenario.
\item A scenario where BSM CP violation is dominated by the chromo-electric dipole moments of down-type quarks.
\end{enumerate}
In each scenario we will compare the reach of searches for EDMs, axionic forces, and axion-induced oscillations of fundamental constants.

\subsection{A  leptoquark scenario}
\label{sec:LQ}

We now consider a BSM model with a $S_1 \in(\bar{3}, 1, 1/3)$  leptoquark (LQ). This scenario induces CP-violating SM-EFT operators which generate both EDMs and oscillations of fundamental constants, allowing us to compare the resulting bounds. The renormalizable couplings of the model are
\begin{align}
	\mathcal{L}_{LQ} = S_1^\gamma \left[\bar{Q}^{c,I}_\gamma y_{LL} \epsilon_{IJ} L^J + \overline{u^c}_R y_{RR} e_R\right] + \text{h.c.}\,,
\end{align}
where $Q$ and $L$ are the quark and lepton doublets respectively, and $\gamma$ is a color index.  We focus on couplings to first-generation quarks and leptons. We have not considered operators without quarks that in principle are also induced in the model as they are less interesting from the present point of view, and we also switch off LQ couplings with two quarks. We assume the leptoquarks to be heavy in order to avoid LHC constraints, which probe masses between $1$ and $2$ TeV \cite{ATLASCollaboration2020Jun,CMSCollaboration2018Nov} for LQs that mainly couple to the first generation, and integrate them out at their threshold scale.  

At $\mu\simeq m_{S_1}$, this leads first of all to a large correction to $\bar \theta$ \cite{deVries:2021sxz,Dekens:2022gha} \begin{equation}\label{thetarenorm}
|\delta \bar \theta| \simeq  \frac{1}{(4\pi)^2} \frac{m_e}{m_u} \mathrm{Im} \left( y_{LL}^*y_{RR} \right)\,.
\end{equation}
 Unless we tune the phase of the Yukawa couplings to very small values by hand, the correction to $\bar \theta$ is unacceptably large. Apart from fine-tuning the tree-level contribution to $\bar \theta$ to cancel the $\delta \bar\theta$, the only other solution is to relax $\bar \theta$ to small values in the infrared through the axion mechanism. 
 
 \begin{figure}[t!]
	\centering
	\includegraphics[width=0.9\linewidth]{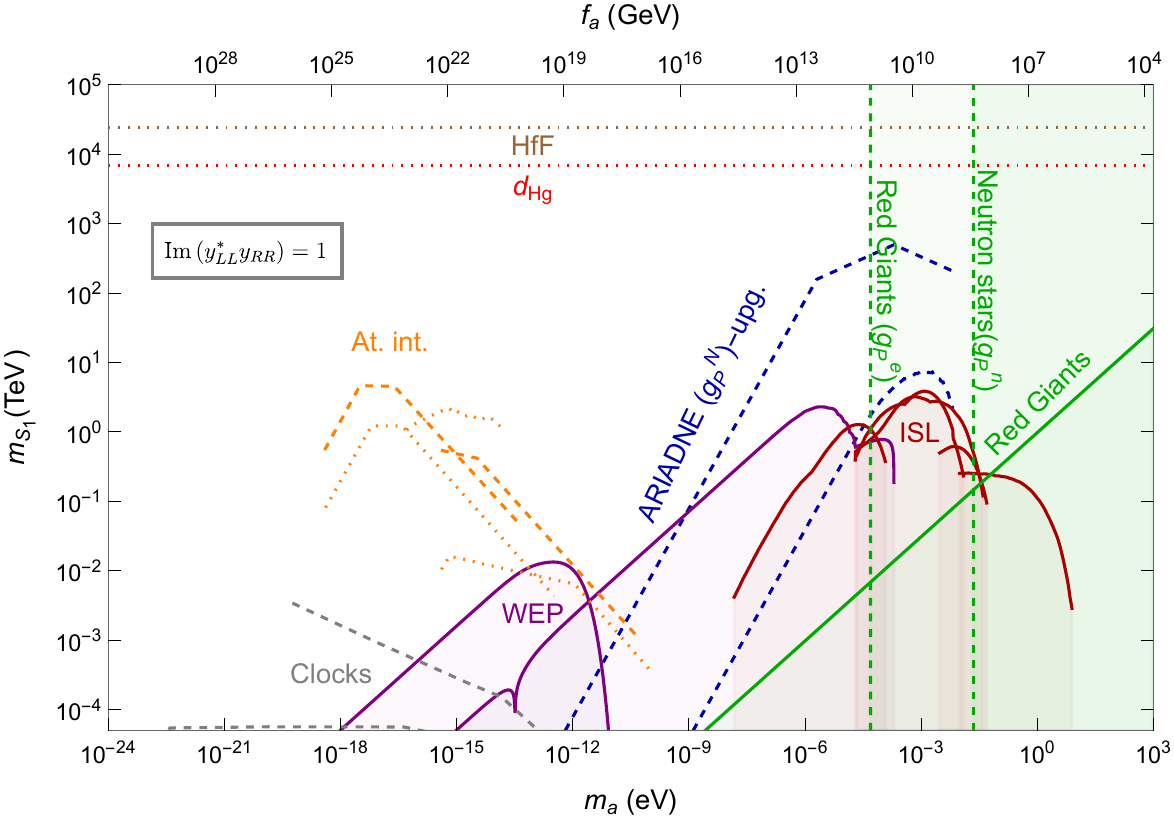}
	 \caption{Constraints on the leptoquark mass as a function of axion mass (or decay constant) as discussed in Sect.~\ref{sec:LQ}, for a scenario with mixing between up-quark and electron. The LQ Yukawa couplings are fixed to $1$ as indicated in the figure. The atomic and nuclear clock bounds are shown in grey, and the prospects for atom interferometry experiments are shown orange (see Fig.~\ref{fig:axphoton} for details); note that the existing bounds are too weak to enter the picture. The reader is referred to Ref.~\cite{Dekens:2022gha} for an explanation of the bounds not discussed here.
	 }
	\label{fig:mLQ}
\end{figure}
 
In addition to renormalizing $\bar \theta$, integrating out LQs leads to several dimension-six SM-EFT operators. Here we are mainly interested in dimension-six electron-quark interactions which, once mapped to LEFT, take the form
\begin{equation}
\mathcal L_{eu} = L^\text{S,RR}_{\substack{eu\\ eeuu}}\, (\bar e_L e_R)(\bar u_L u_R)\,,\qquad L^\text{S,RR}_{\substack{eu\\ eeuu}} = -\frac{1}{2}\frac{y_{LL}^*y_{RR}}{m_{S_1}^2}\,.
\end{equation}
If $L^\text{S,RR}_{\substack{eu\\ eeuu}}$ has a nonzero imaginary part, then the electron-quark interaction violates CP symmetry and contributes to the EDMs of polar molecules such as ThO and HfF \cite{Dekens:2018bci,ACME:2018yjb, Roussy:2022cmp}. The contributions of the above interaction to EDMs are summarized in App.\ \ref{app:edm}. In addition, as worked out in Ref.~\cite{Dekens:2022gha}, the PQ mechanism results in the CP-odd axion-electron coupling
\begin{align}
	\bar{g}_{aee} = \frac{m_*}{2 f_a} \frac{m_\pi^2 \Fp^2}{m_u + m_d}\frac{1}{m_u} \text{Im}\left(L^\text{S,RR}_{\substack{eu\\ eeuu}}\right)\,,
		\label{eq:LQaxel}
\end{align}
and we stress that this interaction depends on the same combination of couplings that enters the correction to $\bar \theta$ in Eq.~\eqref{thetarenorm}. The LQ extension thus nicely illustrates the schematic picture in Fig.~\ref{fig:flowchart}.
In what follows, we will set $\mathrm{Im} \left( y_{LL}^*y_{RR} \right) =1 $ for the couplings to up-quark and electrons in order to illustrate the constraints. Other choices of couplings can simply be obtained by rescaling the value of $m_{S_1}$. 

The upper bounds on $\gae$ shown in Fig.~\ref{fig:gaemodind} can now be understood in terms of the LQ parameters and can be seen to constrain the combination Im$(y_{LL}^* y_{RR})/(f_a m_{S_1}^2)$.  The bounds on the leptoquark mass, as a function of $m_a$ (or $f_a$) are depicted in Fig.~\ref{fig:mLQ}. The projected bounds from the oscillating $\fsc$ probes are several orders of magnitude better than the WEP constraints for small axion masses. The EDM constraints from HfF and d$_\text{Hg}$ measurements are shown as horizontal lines in the plot, while the bound from ThO is slightly weaker than the HfF limit and is thus not shown. These limits are far more stringent than any other bounds across the whole axion mass range.

\textbf{Coupling to muons.} 
If we instead allow the LQ to couple to different generations, the bounds on $m_{S_1}$ can change dramatically. Consider a scenario where the axion couples to muons and up quarks with $\mathcal{O}(1)$ Yukawa couplings and sizable phases, while the other elements of $y_{LL}$ and $y_{RR}$ are set to zero. Again a large threshold correction to $\bar \theta$ is induced which requires an axion solution which leads to a nonzero CP-odd coupling $\bar{g}_{a\mu\mu}$. 

However, in this case the EDM limits are far weaker. The muon EDM is induced when evolving the quark-muon four-fermion interaction to lower energies and we obtain 
\begin{align}
	\left|\frac{d_\mu}{e}\right| \approx \frac{1}{m_{S_1}^2}\frac{m_u}{(4\pi)^2} \left[4 \log\left(\frac{m_{S_1}}{\Lambda_\chi}\right)+\frac{7}{2}\right]+\dots
\end{align}
where $\Lambda_\chi = 2$ GeV is a low-energy scale where perturbative QCD should be matched to chiral EFT. In addition there is a non-perturbative contribution arising from this matching to chiral EFT, which comes with a poorly known QCD matrix element \cite{Dekens:2018pbu}. As the nonperturbative contribution is expected to be of the same size as the perturbative term, we use the latter to estimate the total contribution. The best muon EDM limit arises from considering its contributions to the EDMs of polar molecules \cite{Ema:2021jds, Roussy:2022cmp} which gives approximately
\begin{align}
	|d_\mu| < 7 \times 10^{-21} \, e\,\mathrm{cm}\,,
\end{align}
which is more stringent than the direct limit \cite{Muong-2:2008ebm}. 

\begin{figure}[t!]
 	\centering
 	\includegraphics[width=0.9\linewidth]{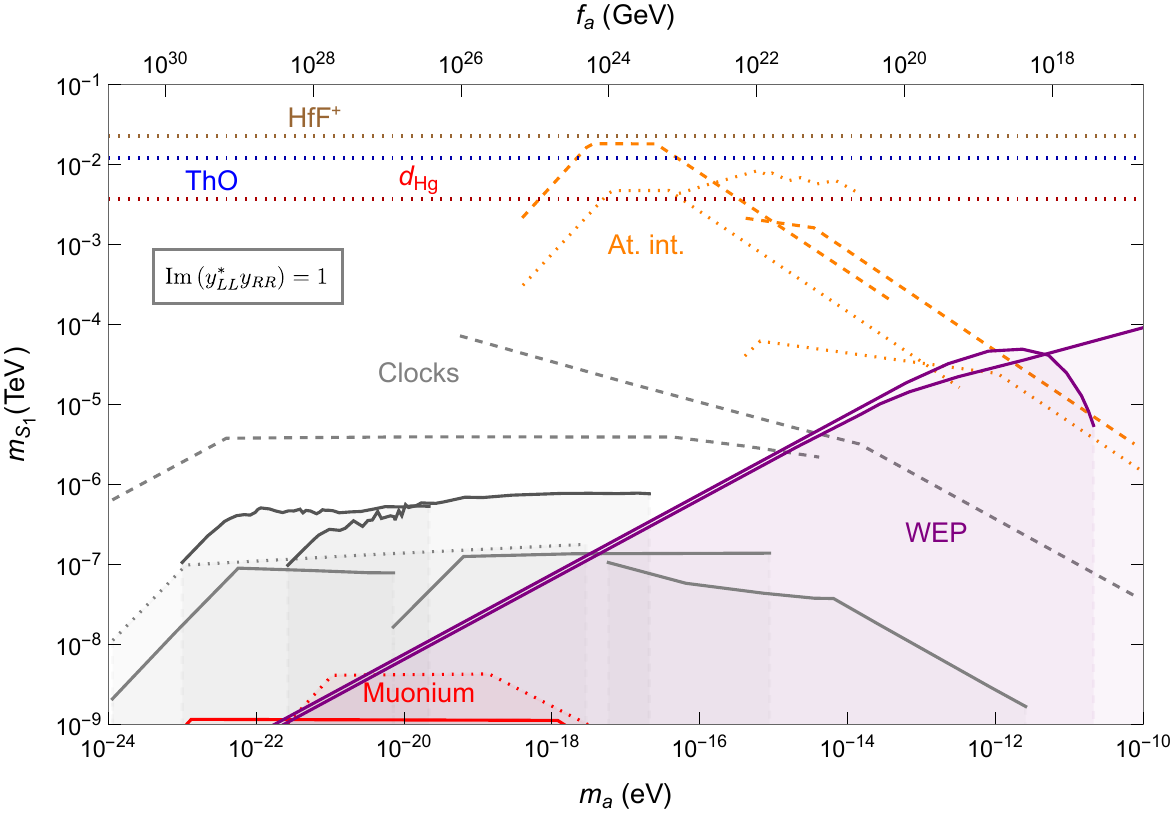}
 	\caption{Constraints on the leptoquark mass as a function of axion mass (or decay constant) as discussed in Sect.~\ref{sec:LQ}, for a scenario with mixing between up-quark and muon. Once again, the Yukawa couplings are set to $1$ as shown in the figure. Unlike Fig.~\ref{fig:mLQ}, the current bounds from clocks do improve the WEP limits for small $m_a$. The lower EDM bounds are from Ref.~\cite{Ema:2021jds}, while the HfF$^+$ bound uses the same prescription but with more recent data~\cite{Roussy:2022cmp}. Rest of the limits and projections can be inferred from Figs.~\ref{fig:axmu} and \ref{fig:mLQ}.}
 	\label{fig:mLQmuon}
\end{figure}

We collect the various limits in Fig.~\ref{fig:mLQmuon}. The EDM limits are now less stringent compared to the electron case but still outperform the present-day searches for fifth forces and time-varying fundamental constants (mainly $\alpha_{\mathrm{em}}(t)$ in this case). Future interferometers could become competitive with the EDM limits in a small window of axion masses, assuming the EDM experiments do not improve. However, we must keep in mind that all bounds for muonic couplings are fairly weak and only probe rather light LQ masses that can already be probed by the LHC experiments. We do not explicitly show the collider constraints, but note that these reach scales of $1-2$ TeV for couplings to muons  \cite{ATLASCollaboration2020Jun}, beyond the mass scales considered in the figure. 

\subsection{Left-right symmetric model}
\label{sec:LRSM}

Left-right symmetric models (LRSM) introduce a new gauge symmetry SU(2)$_R~\times$U(1)~$_{B-L}$ in addition to the SU(3)$_C$ and SU(2)$_L$ symmetries of the SM. This gauge group is broken to U(1)$_Y$ at an energy scale higher than the electroweak scale, characterised by the vev of a right-handed scalar triplet field, $v_R$, which is related to the mass of the right-handed $W$-boson, $m_{W_R}$~\cite{Dekens:2014jka}. In specific variants of the model with exact parity symmetry, the bare QCD theta term is forbidden seemingly resolving the strong CP problem. However, once parity is spontaneously broken at lower energy scales new contributions to $\bar \theta$ are induced, which are significant \cite{Maiezza:2014ala}. These corrections again point towards a PQ mechanism as an attractive solution.  

Within the mLRSM, additional dimension-six operators are induced that violate CP. In particular, the exchange of a $W_R$ leads to a CP-odd four-quark operator that couples left- and right-handed quarks and is sometimes called the four-quark left-right (FQLR) operator \cite{deVries:2012ab}:
\begin{align}
\mathcal L_{6,\,\mathrm{FQLR}} = i C_{\mathrm{FQLR}}\left( \bar u_L \gamma^\mu d_L \,  \bar d_R \gamma_\mu u_R - \bar d_L \gamma^\mu u_L \,  \bar u_R \gamma_\mu d_R\right)\,,
\end{align}
where
\begin{align}
C_{\mathrm{FQLR}} =  |V_{ud}|^2 \frac{g_R^2}{m_{W_R}^2} \frac{\xi \sin \alpha}{1+\xi^2}\,,
\end{align}
in terms of the $SU(2)_R$ gauge coupling, $g_R =g$, a ratio of vacuum expectation values, $\xi$, which determines the amount of $W_L$-$W_R$ mixing, and a spontaneous CP-violating phase, $ \alpha$. The FQLR operator leads to a very rich EDM phenomenology of nucleons, nuclei, and diamagnetic atoms which has been worked out in detail in the literature \cite{An:2009zh, deVries:2012ab, Dekens:2021bro}. In particular, the operator contributes to the nucleon EDMs, as well as pion-nucleon couplings which can induce nuclear EDMs, see App.\ \ref{app:edm} for details.

The combination of the FQLR operator with the PQ mechanism leads to CP-odd axion interactions with hadrons. In particular, it gives rise to the following  axion-nucleon coupling \cite{Dekens:2022gha}
 \begin{align}
	\bar g_{aNN}^{(0)}\simeq \left(0.11 \text{ GeV}^2\right) \frac{\Fp}{f_a} \,C_{\mathrm{FQLR}}\,,
	\end{align}
	in agreement with the NDA expectations in Table~\ref{tab:LEFTNDA} for the $L_{qq}$ operators. 
Within the model, this coupling dominates axionic fifth forces. In addition, in the presence of axion DM it leads to time-varying nucleon masses which can be probed through their impact on hyperfine transitions. However, somewhat more stringent constraints can be set through the induced axion-pion interactions 
\begin{align}
	\bar g_{a\pi\pi}^{(0)}\simeq -(0.45 \text{ GeV})^2\frac{F_\pi^2}{f_a} \, C_{\mathrm{FQLR}} \,.
\end{align} 
This result again agrees with the NDA estimate which estimated $|\bar g_{a\pi\pi}^{(0)}| = \mathcal O(F_\pi^2 \Lambda_\chi^2/f_a\,C_{\mathrm{FQLR}})$. 
These couplings lead to changes in the fine-structure constant through Eq.~\eqref{gap_renorm} with a scaling $\bar g_{a\gamma\gamma} = (\alpha/(4\pi))(\bar g_{a\pi\pi}^{(0)}/m_\pi^2)$ and are thus enhanced by two powers in the chiral power counting, $\Lambda_\chi^2/m_\pi^2$, over the direct contributions given in Table~\ref{tab:LEFTNDA}.

\begin{figure}[t!]
	\centering
	\includegraphics[width=0.9\linewidth]{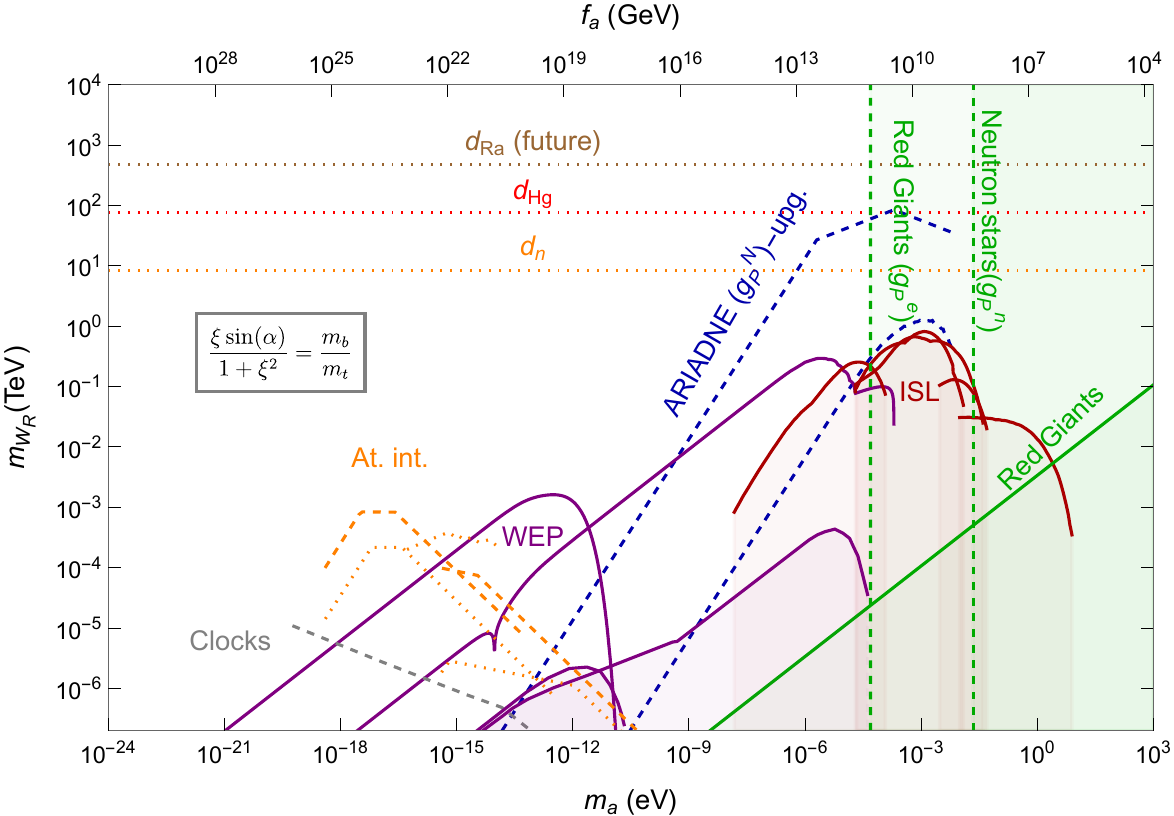}
	\caption{Constraints on the right-handed $W$-boson mass as discussed in Sect.~\ref{sec:LRSM}, with the parameters related to $W_L-W_R$ mixing fixed as indicated in the figure. The atomic clock and interferometer bounds are derived essentially from $\gap$ limits in Fig.~\ref{fig:axphoton}. Once again, the current clock bounds are too weak to be shown.}
	\label{fig:axnuclam}
\end{figure}

We collect all relevant limits in Fig.~\ref{fig:axnuclam} where we present the constraints on $m_{W_R}$ as a function of the axion mass. To generate these plots we have set $\xi/(1+\xi^2) = m_b/m_t$ \cite{Senjanovic:2015yea,Dekens:2021bro} and $\sin \alpha =1$. We observe that EDM limits are much more stringent than current searches for fifth-forces and oscillating fundamental constants. Collider searches also provide constraints up to $m_{W_R} \gtrsim 5$ TeV (not shown in Fig.~\ref{fig:axnuclam}) for $W_R$ bosons decaying to dijets \cite{ATLAS:2017eqx,CMS:2018mgb}, top and bottom quarks \cite{CMS:2021mux,CMS:2023ldh,ATLAS:2023ibb}, or heavy neutrinos~\cite{ATLAS:2019isd,CMS:2021dzb}. Projected atomic clock experiments are competitive with fifth force experiments for $m_a \simeq 10^{-18}$ eV, but are very far away from EDM limits. Future interferometer experiments could significantly improve on existing axion searches in the same axion mass range, but would still be many orders of magnitude away from EDM limits. Somewhat more promising is the proposed ARIADNE experiment, which does not rely on axions providing the DM abundance, but looks for axion-induced monopole-dipole interactions. The projected constraints could overtake EDM limits for axion masses around $10^{-4}$ eV as already pointed out in Refs.~\cite{Bertolini:2020hjc,Dekens:2022gha}. 

\subsection{Chromo-electric dipole moments} 
\label{sec:CEDM}

As the final example, we consider a model where the CP-violation is dominated by chromo-electric dipole moments (CEDMs). Employing only CEDM terms for the first generation of quarks, the Lagrangian is
\begin{align}
	\mathcal{L}_\text{CEDM} &= L_5^u \bar u_L T^A G_{\mu \nu}^A \sigma^{\mu \nu} u_R + L_5^d \bar d_L T^A G_{\mu \nu}^A \sigma^{\mu\nu} d_R + \text{h.c.}\,,
\end{align}
where $T^A$ are the SU(3)$_C$ generators and we will assume that the Wilson coefficients scale as $L_5^q \sim m_q/\Lambda^2$. For concreteness, we turn on only the down-quark CEDM. Similar to the FQLR operator, this interaction induces the nucleon EDMs and pion-nucleon couplings which can generate nuclear EDMs, see App.\ \ref{app:edm} for details. In addition, the PQ mechanism then results in an axion-nucleon coupling~\cite{Dekens:2022gha}
\begin{align}
	\gaN^{(0)}\simeq \left(\frac{0.0054 \text{ GeV}^2}{m_d}\right)\frac{1}{f_a} \text{Im}\left(L_5^d\right),
\end{align}
where we show the $m_d$ dependence explicitly as it cancels out with the $m_d$ scaling in $L_5^d$ when extracting the limits on $\Lambda$. Unlike for the FQLR, for the CEDMs no axion-pion-pion interaction is induced at this order. 

\begin{figure}[t!]
	\centering
	\includegraphics[width=0.9\linewidth]{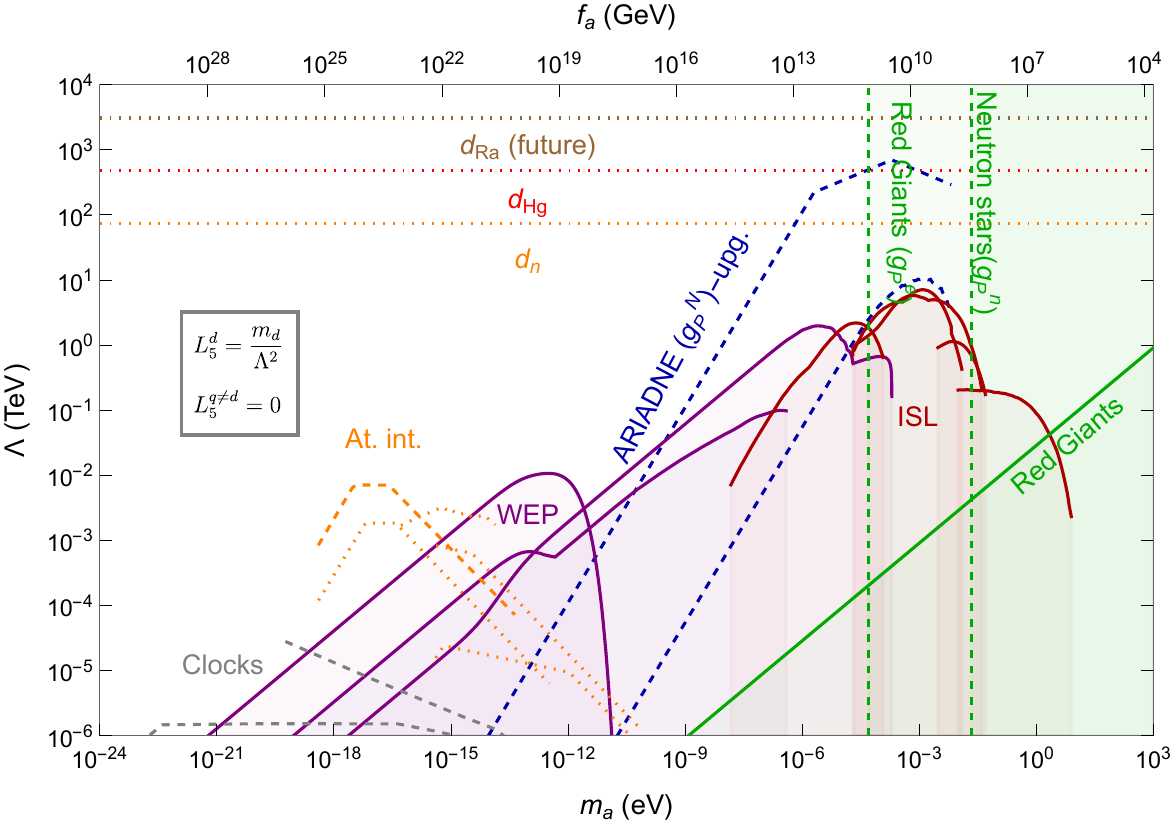}
	\caption{Constraints on the BSM scale $\Lambda$ for the CEDM scenario discussed in Sect.~\ref{sec:CEDM}, where only the down-quark CEDM contribution is considered.}
	\label{fig:cedmplot}
\end{figure}

The bounds on the BSM scale $\Lambda$ in this case are shown in Fig.~\ref{fig:cedmplot}. Like the previous example, the atomic clock and interferometer bounds are rather weak although they do better than WEP bounds for tiny $m_a$ in this case as well. The EDM limits completely dominate across all mass scales and only the future ARIADNE experiment has any hope of competing with these constraints.

\section{Conclusions and discussion}\label{discussion}
Axions can provide a simultaneous explanation of two problems in the SM: the lack of a DM candidate and the absence of strong CP violation. Already within the SM there exist more sources of CP violation and there are good reasons to believe that BSM theories contain additional sources, for instance to account for the universal matter/antimatter asymmetry. In presence of non-$\bar \theta$ sources of CP violation, the Peccei-Quinn mechanism leads to CP-violating interactions between axions and SM fields, and, at lower energies, to CP-violating interactions between axions and hadrons. These interactions were constructed in the framework of the Standard Model EFT in Ref.~\cite{Dekens:2022gha}, which provides a direct connection between the strength and form of CP-odd axion interactions with hadrons and leptons and general underlying sources of CP violation at the quark level.

In this work we go one step further and assume that axions are light and form the DM of our universe. In this case, the CP-odd axion couplings can effectively be interpreted as time-oscillating fundamental constants. For example, the fine-structure constant $\fsc$ and SM particle masses will now oscillate with a frequency set by the axion mass and an amplitude that depends on the strength of the CP-violating sources. The resulting phenomenology is very similar to that of ultralight scalar DM. The oscillations can be probed by precision experiments that, for example, look for variations in atomic transition frequencies. In the first part of this paper, we work out how a variety of CP-violating axion couplings are connected to searches for oscillating fundamental constants. In particular, we demonstrate that loop corrections can be relevant because the different experimental searches have very different sensitivities so that loop suppressions can be overcome. This part of our work is independent of the axionic nature of DM and equally applies to scalar DM.

We subsequently compared fifth-force experiments, test of the weak equivalence principle (WEP), to searches for time-varying fundamental constants. 
In general we find that for very light axions, the latter probes have the potential to significantly improve the limits on CP-odd axion-photon, -electron, -muon, and -nucleon couplings from fifth-force and WEP experiments. At higher masses, however, the bounds set by clock experiments are weaker as the oscillation frequency becomes too large to be detected with considerable precision. This could be partially overcome by future interferometer experiments. We find that the most sensitive searches are related to measurements of optical transitions which can constrain a time-varying fine-structure constant. For example, CPV axion-muon couplings lead to oscillating muon masses and, through loop corrections, to oscillating $\alpha_{\mathrm{em}}$. In this case, searches for the latter outperform more direct limits by 5 orders of magnitude. Similarly, searches for CPV axion-electron interactions through hyperfine transitions in microwave clocks cannot compete with the indirect limits obtained from optical transitions. 

In the second part of the paper, we focus on the specific case of axion DM. In this case, the CP-violating axion-SM interactions are sourced by non-$\bar \theta$ sources of CP violation and can therefore be related to axion-independent CP-violating effects. We consider several scenarios involving such sources of CP violation, both within the Standard Model and beyond. The CPV axion couplings can be expressed in terms of the model parameters in these scenarios, allowing us to see how the constraints from (a lack of) variation in fundamental constants compare to direct limits on extra sources of CP violation from electric dipole moment experiments. While future searches for time-varying fundamental constants can greatly improve limits obtained from fifth-force experiments, we find that, in general, they fall short of EDM sensitivities. That is, mechanisms that would predict, for example, a detectable axion-induced time-varying fine-structure constant within axion models, should have led already to a detection in any of the recent electric dipole moment experiments. This conclusion can be avoided when the axion does not couple to first-generation particles, for example, the EDM limits are significantly less stringent in the case of couplings to muons. Furthermore, EDM limits do not apply for ultralight scalar DM so that the search for time-varying constants remains well motivated. Finally, future prospects, for instance through nuclear clocks and/or quantum sensing \cite{Bass:2023hoi}, have the potential to greatly enhance the search for time-varying constants and it might very well be that such searches provide the best chance to measure CP-violating axion interactions in the future. 

\acknowledgments
We thank Arghavan Safavi Naini for useful discussions and encouragement. 
JdV acknowledges support from the Dutch Research Council (NWO) in the form of a VIDI grant. WD acknowledges support by the U.S. DOE under Grant No. DE-FG02-00ER41132.

\appendix
\section{Contributions of LEFT operators }
\label{app:BunchOfOps}

Here we discuss the contributions of LEFT operators to Eq.~\eqref{axionCPV}. The operators that are expected to give the leading contributions, those unsuppressed by derivatives in $\chi$PT, are listed in Table~\ref{tab:oplist1} \cite{Dekens:2022gha}.
 Here we only consider the operators that give rise to unsuppressed operators in the chiral Lagrangian. The relevant four-quark operators arise from the $(\bar LL)(\bar RR)$, $(\bar LR)(\bar LR)$, and $(\bar LR)(\bar RL)$ classes. The unsuppressed semi-leptonic operators are generated by the $(\bar LR)(\bar LR)$ class, while the dipole and Weinberg operators arise from the $(\bar LR) X$ and $X^3$ classes.

We employ NDA to estimate the contributions of these interactions to the axion couplings in Eq.~\eqref{axionCPV} and list the results in Table~\ref{tab:LEFTNDA}. All estimates are based on NDA, with the exception of the contribution to $\bar g_{a\pi\pi}^{(0)}$ from the quark CEDMs, shown in blue. This contribution would be estimated by $ \Lambda_\chi^2 \frac{F_\pi}{f_a}$ if one would simply apply NDA. However, explicit calculation shows \cite{Dekens:2022gha} that these leading terms vanish after aligning the vacuum, so that the first contributions are suppressed by a factor of $m_q/\Lambda_\chi$ as shown in the table.

\begin{table}[t!]
\vspace{-0.75cm}
\begin{adjustbox}{width=0.5\textwidth,center}
\begin{minipage}[t]{3cm}
\renewcommand{\arraystretch}{1.51}
\small
\begin{align*}
\begin{array}[t]{c|c}
\multicolumn{2}{c}{\boldsymbol{(\overline L R ) X+\hc}} \\
\hline
\blue{\O_{u \gamma} }& \bar u_{Lp}   \sigma^{\mu \nu}  u_{Rr}\, F_{\mu \nu}   \\
\blue{\O_{d \gamma} }& \bar d_{Lp}  \sigma^{\mu \nu} d_{Rr}\, F_{\mu \nu}  \\
\O_{u G} & \bar u_{Lp}   \sigma^{\mu \nu}  T^A u_{Rr}\,  G_{\mu \nu}^A  \\
\O_{d G} & \bar d_{Lp}   \sigma^{\mu \nu} T^A d_{Rr}\,  G_{\mu \nu}^A \\
\end{array}
\end{align*}
\end{minipage}
\begin{minipage}[t]{3cm}
\renewcommand{\arraystretch}{1.51}
\small
\begin{align*}
\begin{array}[t]{c|c}
\multicolumn{2}{c}{\boldsymbol{X^3}} \\
\hline
\O_{\widetilde G} & f^{ABC} \widetilde G_\mu^{A\nu} G_\nu^{B\rho} G_\rho^{C\mu}   \\
\end{array}
\end{align*}
\end{minipage}
\end{adjustbox}
\mbox{}\\[-0.75cm]
\begin{adjustbox}{width=1.05\textwidth,center}
\begin{minipage}[t]{3cm}
\renewcommand{\arraystretch}{1.51}
\small
\begin{align*}
\begin{array}[t]{c|c}
\multicolumn{2}{c}{\boldsymbol{(\overline L L)(\overline L  L)}} \\
\hline
\blue{\op{\nu u}{V}{LL}  }     & (\bar \nu_{Lp} \gamma^\mu \nu_{Lr}) (\bar u_{Ls}  \gamma_\mu u_{Lt})  \\
\blue{\op{\nu d}{V}{LL}   }    & (\bar \nu_{Lp} \gamma^\mu \nu_{Lr})(\bar d_{Ls} \gamma_\mu d_{Lt})     \\
\blue{\op{eu}{V}{LL}  }    & (\bar e_{Lp}  \gamma^\mu e_{Lr})(\bar u_{Ls} \gamma_\mu u_{Lt})   \\
\blue{\op{ed}{V}{LL}   }    & (\bar e_{Lp}  \gamma^\mu e_{Lr})(\bar d_{Ls} \gamma_\mu d_{Lt})  \\
\blue{\op{\nu edu}{V}{LL} }     & (\bar \nu_{Lp} \gamma^\mu e_{Lr}) (\bar d_{Ls} \gamma_\mu u_{Lt})  + \hc   \\
[-0.5cm]
\end{array}
\end{align*}
\renewcommand{\arraystretch}{1.51}
\small
\begin{align*}
\begin{array}[t]{c|c}
\multicolumn{2}{c}{\boldsymbol{(\overline R  R)(\overline R R)}} \\
\hline
\blue{\op{eu}{V}{RR}    }   & (\bar e_{Rp}  \gamma^\mu e_{Rr})(\bar u_{Rs} \gamma_\mu u_{Rt})   \\
\blue{\op{ed}{V}{RR}   }  & (\bar e_{Rp} \gamma^\mu e_{Rr})  (\bar d_{Rs} \gamma_\mu d_{Rt})   \\
\end{array}
\end{align*}
\end{minipage}

\begin{minipage}[t]{3cm}
\renewcommand{\arraystretch}{1.51}
\small
\begin{align*}
\begin{array}[t]{c|c}
\multicolumn{2}{c}{\boldsymbol{(\overline L  L)(\overline R  R)}} \\
\hline
\blue{\op{\nu u}{V}{LR}   }      & (\bar \nu_{Lp} \gamma^\mu \nu_{Lr})(\bar u_{Rs}  \gamma_\mu u_{Rt})    \\
\blue{\op{\nu d}{V}{LR}    }     & (\bar \nu_{Lp} \gamma^\mu \nu_{Lr})(\bar d_{Rs} \gamma_\mu d_{Rt})   \\
\blue{\op{eu}{V}{LR}   }     & (\bar e_{Lp}  \gamma^\mu e_{Lr})(\bar u_{Rs} \gamma_\mu u_{Rt})   \\
\blue{\op{ed}{V}{LR}     }   & (\bar e_{Lp}  \gamma^\mu e_{Lr})(\bar d_{Rs} \gamma_\mu d_{Rt})   \\
\blue{\op{ue}{V}{LR}  }      & (\bar u_{Lp} \gamma^\mu u_{Lr})(\bar e_{Rs}  \gamma_\mu e_{Rt})   \\
\blue{\op{de}{V}{LR}     }    & (\bar d_{Lp} \gamma^\mu d_{Lr}) (\bar e_{Rs} \gamma_\mu e_{Rt})   \\
\blue{\op{\nu edu}{V}{LR}  }      & (\bar \nu_{Lp} \gamma^\mu e_{Lr})(\bar d_{Rs} \gamma_\mu u_{Rt})  +\hc \\
\op{uu}{V1}{LR}        & (\bar u_{Lp} \gamma^\mu u_{Lr})(\bar u_{Rs} \gamma_\mu u_{Rt})   \\
\op{uu}{V8}{LR}       & (\bar u_{Lp} \gamma^\mu T^A u_{Lr})(\bar u_{Rs} \gamma_\mu T^A u_{Rt})    \\ 
\op{ud}{V1}{LR}       & (\bar u_{Lp} \gamma^\mu u_{Lr}) (\bar d_{Rs} \gamma_\mu d_{Rt})  \\
\op{ud}{V8}{LR}       & (\bar u_{Lp} \gamma^\mu T^A u_{Lr})  (\bar d_{Rs} \gamma_\mu T^A d_{Rt})  \\
\op{du}{V1}{LR}       & (\bar d_{Lp} \gamma^\mu d_{Lr})(\bar u_{Rs} \gamma_\mu u_{Rt})   \\
\op{du}{V8}{LR}       & (\bar d_{Lp} \gamma^\mu T^A d_{Lr})(\bar u_{Rs} \gamma_\mu T^A u_{Rt}) \\
\op{dd}{V1}{LR}      & (\bar d_{Lp} \gamma^\mu d_{Lr})(\bar d_{Rs} \gamma_\mu d_{Rt})  \\
\op{dd}{V8}{LR}   & (\bar d_{Lp} \gamma^\mu T^A d_{Lr})(\bar d_{Rs} \gamma_\mu T^A d_{Rt}) \\
\op{uddu}{V1}{LR}   & (\bar u_{Lp} \gamma^\mu d_{Lr})(\bar d_{Rs} \gamma_\mu u_{Rt})  + \hc  \\
\op{uddu}{V8}{LR}      & (\bar u_{Lp} \gamma^\mu T^A d_{Lr})(\bar d_{Rs} \gamma_\mu T^A  u_{Rt})  + \hc \\
\end{array}
\end{align*}
\end{minipage}

\begin{minipage}[t]{3cm}
\renewcommand{\arraystretch}{1.51}
\small
\begin{align*}
\begin{array}[t]{c|c}
\multicolumn{2}{c}{\boldsymbol{(\overline L R)(\overline L R)+\hc}} \\
\hline
\blue{\op{eu}{S}{RR} } & (\bar e_{Lp}   e_{Rr}) (\bar u_{Ls} u_{Rt})   \\
\blue{\op{eu}{T}{RR} }& (\bar e_{Lp}   \sigma^{\mu \nu}   e_{Rr}) (\bar u_{Ls}  \sigma_{\mu \nu}  u_{Rt})  \\
\blue{\op{ed}{S}{RR} } & (\bar e_{Lp} e_{Rr})(\bar d_{Ls} d_{Rt})  \\
\blue{\op{ed}{T}{RR} }& (\bar e_{Lp} \sigma^{\mu \nu} e_{Rr}) (\bar d_{Ls} \sigma_{\mu \nu} d_{Rt})   \\
\blue{\op{\nu edu}{S}{RR}} & (\bar   \nu_{Lp} e_{Rr})  (\bar d_{Ls} u_{Rt} ) \\
\blue{\op{\nu edu}{T}{RR} }&  (\bar  \nu_{Lp}  \sigma^{\mu \nu} e_{Rr} )  (\bar  d_{Ls}  \sigma_{\mu \nu} u_{Rt} )   \\
\op{uu}{S1}{RR}  & (\bar u_{Lp}   u_{Rr}) (\bar u_{Ls} u_{Rt})  \\
\op{uu}{S8}{RR}   & (\bar u_{Lp}   T^A u_{Rr}) (\bar u_{Ls} T^A u_{Rt})  \\
\op{ud}{S1}{RR}   & (\bar u_{Lp} u_{Rr})  (\bar d_{Ls} d_{Rt})   \\
\op{ud}{S8}{RR}  & (\bar u_{Lp} T^A u_{Rr})  (\bar d_{Ls} T^A d_{Rt})  \\
\op{dd}{S1}{RR}   & (\bar d_{Lp} d_{Rr}) (\bar d_{Ls} d_{Rt}) \\
\op{dd}{S8}{RR}  & (\bar d_{Lp} T^A d_{Rr}) (\bar d_{Ls} T^A d_{Rt})  \\
\op{uddu}{S1}{RR} &  (\bar u_{Lp} d_{Rr}) (\bar d_{Ls}  u_{Rt})   \\
\op{uddu}{S8}{RR}  &  (\bar u_{Lp} T^A d_{Rr}) (\bar d_{Ls}  T^A u_{Rt})  \\[-0.5cm]
\end{array}
\end{align*}
\renewcommand{\arraystretch}{1.51}
\small
\begin{align*}
\begin{array}[t]{c|c}
\multicolumn{2}{c}{\boldsymbol{(\overline L R)(\overline R L) +\hc}} \\
\hline
\blue{\op{eu}{S}{RL} } & (\bar e_{Lp} e_{Rr}) (\bar u_{Rs}  u_{Lt})  \\
\blue{\op{ed}{S}{RL}} & (\bar e_{Lp} e_{Rr}) (\bar d_{Rs} d_{Lt}) \\
\blue{\op{\nu edu}{S}{RL} } & (\bar \nu_{Lp} e_{Rr}) (\bar d_{Rs}  u_{Lt})  \\
\end{array}
\end{align*}
\end{minipage}
\end{adjustbox}
\setlength{\abovecaptionskip}{0.15cm}
\caption{The $B$- and $L$-conserving operators of the LEFT of dimension five and six that contribute to CP-violating effects in the meson sector at leading order. Only the hadronic operators that contribute to the non-derivative meson interactions and the semi-leptonic operators that can be written as external sources (shown in blue) are listed.}
\label{tab:oplist1}
\end{table}

\renewcommand{\arraystretch}{1.8}
\begin{table}[t!]
\centering
\begin{tabular}{|c |c cc|cc|}
\hline
 &
 $L_{q\gamma}$ & $ L_{qG} $& $ L_{\tilde G}$ &  $L_{qq}$ &  $L_{q\ell}$
 \\\hline
$\bar g_{aNN}^{(0)}$& $e F_\pi \frac{F_\pi}{f_a}$ &  $ \Lambda_\chi \frac{F_\pi}{f_a}$ & $ \Lambda_\chi^2 \frac{F_\pi}{f_a}$ &  $ F_\pi \Lambda_\chi \frac{F_\pi}{f_a}$ & $-$\\
$\bar g_{a\ell\ell}^{(0)}$ & $-$& $-$& $-$& $-$ & $F_\pi \Lambda_\chi \frac{F_\pi}{f_a}$\\
$\bar g_{a\pi\pi}^{(0)}$ & $e F_\pi\Lambda_\chi \frac{F_\pi}{f_a}$ &  ${\color{blue} \Lambda_\chi^2\frac{F_\pi}{f_a}  \left(\frac{m_q}{\Lambda_\chi}\right)}$ & $ \Lambda_\chi^2 m_q\frac{F_\pi}{f_a}$ &  $ F_\pi \Lambda_\chi^2 \frac{F_\pi}{f_a}$ & $-$\\
$\gap^{(0)}$ & $\frac{e}{4\pi}\frac{F_\pi}{f_a}$ & $\frac{\alpha}{4\pi} \frac{F_\pi}{f_a}$ & $\alpha F_\pi \frac{F_\pi}{f_a} $& $\frac{\alpha}{4\pi}F_\pi \frac{F_\pi}{f_a} $ &  $\frac{\alpha}{4\pi}\frac{m_\ell}{\Lambda_\chi}F_\pi\frac{F_\pi}{f_a} $\\
\hline
\end{tabular}
\caption{NDA estimates of the direct contributions from LEFT operators to the axion couplings in Eq.~\eqref{axionCPV}. Here $q$ denotes $q=\{u,d\}$ in the Wilson coefficients $L_{q\gamma}$ and $L_{qG}$. $L_{qq}$ and $L_{q\ell}$ denote flavor-diagonal hadronic and semi-leptonic four-fermion operators in Table~\ref{tab:oplist1} that give rise to chirally unsuppressed interactions in the chiral Lagrangian, see the text for details. The contribution of $L_{qG}$ to $\bar g_{a\pi\pi}^{(0)}$, indicated in blue, is smaller than its naive NDA estimate as explained in the text.
}
\label{tab:LEFTNDA}
\end{table}

\section{Contributions to EDMs}\label{app:edm}

The CP-odd electron-nucleon and pion-nucleon interactions discussed in Section \ref{sec:apps} contribute to the EDMs of nucleons, nuclei, atoms, and molecules. Here we summarize the relevant input needed to estimate these effects.

\subsection{EDMs of polar molecules}
The semi-leptonic operators considered in Section \ref{sec:LQ}, contribute to CP-odd scalar and pseudoscalar interactions between nucleons and electrons \cite{Dekens:2018bci},
\begin{eqnarray}
C_S^{(0)} &= -v_H^2\frac{\sigma_{\pi N}}{m_u+m_d}{\rm Im}\left[L_{\substack{eu\\eeuu}}^{\rm S,RR}
\right] ,\qquad 
C_S^{(1)} &= -v_H^2\frac{1}{2}\frac{\delta m_N}{m_d-m_u}{\rm Im}\left[L_{\substack{eu\\eeuu}}^{\rm S,RR}\right]\,,\nn
C_P^{(0)} &= v_H^2\frac{m_N B(D-3F)}{3m_\eta^2}{\rm Im}\left[L_{\substack{eu\\eeuu}}^{\rm S,RR}\right]\,,\qquad
C_P^{(1)} &= -v_H^2\frac{m_N B g_A}{m_\pi^2}{\rm Im}\left[L_{\substack{eu\\eeuu}}^{\rm S,RR}\right]\,,
\end{eqnarray}
where $v_H$ is the vev of the Higgs field, at tree level $v_H^2 =\sqrt{2}G_F \simeq (246\, {\rm GeV})^2$.
Furthermore, $g_A = D+F$ is the axial charge of the nucleon, $\delta m_N = (m_n-m_p)_{QCD}$ is the strong nucleon mass splitting, while the nucleon sigma terms are given by $\sigma_q = m_q\frac{\partial \Delta m_N}{\partial m_q}$, where $\Delta m_N = \frac{m_n+m_p}{2}$, and $\sigma_{\pi N} = \sigma_u+\sigma_d$. 
The input for these hadronic matrix elements can be summarized as \cite{Airapetian:2006vy,Hoferichter:2015dsa,Abdel-Rehim:2016won,Borsanyi:2014jba,Brantley:2016our} 
\begin{eqnarray}\label{sigma}
\sigma_{\pi N} &=& (59.1 \pm 3.5)\,\mathrm{MeV}\ ,\qquad 
\delta m_N = (2.32\pm 0.17)\,{\rm MeV}
 \, ,\nn
\delta m_N &=&( 2.32\pm0.17)\, \mathrm{MeV}\, ,\qquad
  g_A = 1.27\pm0.002\,.
\end{eqnarray}

The scalar nucleon-electron couplings, $C^{(0,1)}_S$, contribute to CP-odd effects in polar molecules \cite{doi:10.1063/1.4968597,Fleig:2017mls,PhysRevA.93.042507}
\begin{eqnarray}
\omega_{\text{HfF}} &=&+(32.0\pm1.3)(\mathrm{mrad}/\mathrm{s})\left(\frac{C_S }{10^{-7}}\right)\,,\\
\omega_{\text{ThO}} &=&+(181.6\pm7.3)(\mathrm{mrad}/\mathrm{s})\left(\frac{C_S }{10^{-7}}\right)\, ,
\label{eq:Molecules}
\end{eqnarray}
in terms of $C_S = C_S^{(0)}+\frac{Z-N}{Z+N} C_S^{(1)}$ where $Z$ and $N$ correspond to the number of protons and neutrons, respectively, of the heaviest atom of the molecule. 

\subsection{EDMs of nucleons, nuclei, and atoms}
In addition to the effects in polar molecules, the semileptonic interactions and the isoscalar and isovector pion-nucleon couplings, $\bar g_{0,1}$, induce EDMs of nucleons, nuclei, and diamagnetic atoms. 
For the FQLR and CEDM operators discussed in Sections \ref{sec:LRSM} and \ref{sec:CEDM}, the relevant pion-nucleon couplings are given by \cite{Dekens:2022gha},
\begin{eqnarray}
\bar g_0 &=&-\frac{\bar B}{2F_\pi B}\frac{\delta m_N}{\bar m \epsilon}{\rm Im}\left(L_5^d\right)\,,\nn
\bar g_1&=& \frac{\bar B}{F_\pi B}\frac{\Delta m_N}{\bar m}{\rm Im}\left(L_5^d\right)-0.62 \,{\rm GeV}^2\,  C_{\rm FQLR} \,,
\end{eqnarray}
where $\bar m = \frac{m_u+m_d}{2}$,  $\bar m \epsilon = \frac{m_d-m_u}{2}$, and $\bar B/B$ is a ratio of matrix elements of the chromo-magnetic operator and the quark condensate. This ratio has been estimated using QCD sum rules \cite{Pospelov:2001ys} as well as through a relation to deep inelastic scattering \cite{ Seng:2018wwp}. Here we follow Ref.~\cite{Seng:2018wwp} and use $\bar B/B \simeq 0.4\,\mathrm{GeV}^2/g_s(2\,\mathrm{GeV})$ \cite{Belyaev:1982sa, Seng:2018wwp} with $g_s(2\,\mathrm{GeV})\simeq 1.85$.
In the above expressions for $\bar g_{0,1}$ we neglected possible `direct' contributions to the pion-nucleon couplings, see Ref.~\cite{Dekens:2022gha} for details.

The nucleon EDMs can now be estimated in terms of the nucleon-pion couplings through \cite{Seng:2014pba},
\begin{eqnarray}
d_n &=& - \frac{e g_A }{8 \pi ^2 F_\pi} \left[\bar g_0 \left( \log \frac{m^2_\pi}{m_N^2} - \frac{\pi m_\pi}{2 m_N} \right)  + \frac{\bar g_1}{4 } \left( \kappa_1 - \kappa_0\right) \frac{m^2_\pi}{m_N^2} \log \frac{m^2_\pi}{m_N^2}  \right] \ ,
\label{eq:dn}
\\
d_p &=&  \frac{e g_A}{8 \pi^2 F_\pi} \Bigg[  \bar g_0\left( \log \frac{m^2_\pi}{m_N^2} - \frac{2 \pi m_\pi}{m_N} \right)  
-\frac{\bar g_1}{4 } \left(  \frac{2 \pi m_\pi}{m_N} + \left( 5/2 + \kappa_1 + \kappa_0\right) \frac{m^2_\pi}{m_N^2} \log \frac{m^2_\pi}{m_N^2}  \right)  \Bigg]
\ ,\nonumber
\label{eq:dp}
\end{eqnarray}
where $\kappa_1 = 3.7$ and $\kappa_0 = -0.12$ are related to the nucleon magnetic moments. 
Here we set the renormalization scale to the nucleon mass $m_N$ in order to estimate the EDMs as function of pion-nucleon couplings, but note that these EDMs in principle also receive direct contributions from the FQLR and CEDM operators. Such contributions depend on poorly controlled matrix elements and we neglect them here, see the discussion in e.g.\  Ref.\ \cite{Dekens:2022gha} for more details.

\begin{table}[t]
\small\center
\renewcommand{\arraystretch}{1.2}
$\begin{array}{ccc|ccc}
\multicolumn{3}{c|}{{\rm neutron}\, {\rm and }\,{\rm atoms }  \,(e \, {\rm cm})} &\multicolumn{2}{c}{{\rm Molecules}  \,(\mathrm{mrad}/\mathrm{s}) }   \\\hline
d_n & d_{\rm Hg} & d_{\rm Ra} &\omega_{\text{HfF}} & \omega_{\text{ThO}}\\\hline 
1.8 \cdot 10^{-26} &6.3\cdot 10^{-30} & 1.2\cdot 10^{-23}&
  0.14
&
1.3\\
  \end{array}$
\caption{Current experimental limits (at 90$\%$ C.L.) from measurements on the neutron \cite{Abel:2020pzs}, $^{199}$Hg \cite{Graner:2016ses}, $^{225}$Ra \cite{Bishof:2016uqx}, YbF
\cite{Hudson:2011zz}, HfF \cite{Roussy:2022cmp}, and ThO \cite{ACME:2018yjb}.}\label{tab:expt}
\end{table}

The Hg EDM can now be written as
\cite{Engel:2013lsa,Fleig:2018bsf,Dzuba:2009kn,Latha:2009nq,Yamanaka:2017mef,Dmitriev:2003sc},
\begin{eqnarray}\label{dHg} 
d_{\rm Hg}&=& -(2.1\pm0.5)\Ex{-4}\bigg[(1.9\pm0.1)d_n +(0.20\pm 0.06)d_p\nn
&&+\bigg(0.13^{+0.5}_{-0.07}\,\bar g_0 +
0.25^{+0.89}_{-0.63}\,\bar g_1\bigg)e\, {\rm fm}\bigg]\nn
&& - \left[ (0.028\pm0.006) C_S- \frac{1}{3}(3.6\pm0.4) \left(\frac{Z\alpha}{5 m_N R}C_P\right)\right]\cdot
10^{-20}\, e\,\mathrm{cm}\,,
\end{eqnarray}
in terms of the nuclear radius $R\simeq 1.2\, A^{1/3}$ fm, and
$C_{P} = (C_{P}^{(n)}\langle \vec \sigma_n\rangle  +C_{P}^{(p)}\langle \vec \sigma_p\rangle
)/(\langle \vec \sigma_n\rangle +\langle \vec \sigma_p\rangle )$. Here we defined the neutron and proton couplings as $C_{P}^{(n,p)}=C_{P}^{(0)}\mp C_{P}^{(1)}$. For $^{199}$Hg
we use the values \cite{Yanase:2018qqq} 
\begin{eqnarray}
\langle \vec \sigma_n\rangle= -0.3249\pm0.0515\,,\qquad \langle \vec \sigma_p\rangle = 0.0031\pm 0.0118\,.
\end{eqnarray}
The atomic EDM of Ra is dominated by nuclear CP violation due to octopole-deformation of the nucleus, which somewhat simplifies the relevant expression
\cite{Engel:2013lsa,Dobaczewski:2018nim}
\begin{eqnarray}\label{dRa}
d_{\mathrm{Ra}} &=& (7.7\cdot 10^{-4})\cdot\left[(2.5\pm 7.5)\,\bar g_0 - (65 \pm 40)\,\bar g_1\right]e\, {\rm fm}\,.
\end{eqnarray}
The pion-nucleon couplings give the largest contributions to the Ra and Hg EDMs as long as $\bar g_{0,1}$ receive contributions at leading order, which is the case for the FQLR and CEDM operators under consideration here. The current best limits are collected in Table~\ref{tab:expt}.

\bibliographystyle{JHEP_mod}
\bibliography{bibfile}

\end{document}